\documentclass{article}
\usepackage{color}
\usepackage{graphicx}
\usepackage{amsmath, amsthm}
\usepackage{mathtools}
\usepackage{natbib}
\usepackage{url}
\usepackage{soulutf8}
\RequirePackage[colorlinks,citecolor=blue,urlcolor=blue]{hyperref}
\usepackage{xcolor}
\usepackage{chngcntr}
\usepackage{subfig}
\usepackage{hyperref}
\usepackage{algorithm}
\usepackage[noend]{algpseudocode}
\usepackage{setspace}
\newtheorem{theorem}{Theorem}
\newtheorem{corollary}{Corollary}
\usepackage{xfrac}
\definecolor{forest}{rgb}{0.133,0.545,0.133}
\usepackage{multirow}
\usepackage{amsfonts}

\evensidemargin=\oddsidemargin
\usepackage{etoolbox}

\newif\ifabbreviation
\pretocmd{\thebibliography}{\abbreviationfalse}{}{}
\AtBeginDocument{\abbreviationtrue}

\begin{document}
	\newcommand{\bb}{\boldsymbol{\beta}}

	\title{Posterior Ramifications of Prior Dependence Structures}


\author{Luke Hagar\footnote{Luke Hagar is the corresponding author and may be contacted at \url{luke.hagar@mail.mcgill.ca}.} \hspace{35pt} Nathaniel T. Stevens$^{\dagger}$ \bigskip \\ 
 $^*$\textit{Department of Epidemiology, Biostatistics \& Occupational Health, McGill University} \\ $^{\dagger}$\textit{Department of Statistics \& Actuarial Science, University of Waterloo}}

	\date{}

	\maketitle

	\begin{abstract}

Prior elicitation methods for Bayesian analyses transfigure prior information into quantifiable prior distributions. Recently, methods that leverage copulas have been proposed to accommodate more flexible dependence structures when eliciting multivariate priors. We show that \color{black}{the} posterior cannot retain many of these flexible prior dependence structures in large-sample settings, \color{black}{and we emphasize that it is our responsibility as statisticians to communicate this to practitioners}. \color{black}We therefore overview objectives for prior specification \color{black}{that guide conversations between statisticians and practitioners to promote alignment between the flexibility in the prior dependence structure and} \color{black}{the} objectives for posterior analysis. Because correctly specifying the dependence structure a priori can be difficult, we consider how the choice of prior copula impacts the posterior distribution in terms of asymptotic convergence of the posterior mode. Our resulting recommendations \color{black}{clarify when it is useful to elicit intricate prior dependence structures and when it is not.}
\end{abstract}

		\bigskip


	\maketitle

	\baselineskip=19.5pt


     \section{Introduction}

    \subsection{Overview of Prior Elicitation}

   Statistical methods leverage data to infer properties about an unobservable, possibly multivariate parameter $\boldsymbol{\theta} = (\theta_1, ..., \theta_d)$. Bayesian methods for statistical inference (see e.g., \citep{gelman2013bayesian}) require the specification of a prior distribution for the parameter $\boldsymbol{\theta}$, denoted by $p(\boldsymbol{\theta})$. This distribution characterizes the beliefs about $\boldsymbol{\theta}$ prior to observing any data. For a particular statistical model, $L(\boldsymbol{\theta};\hspace{1pt} \boldsymbol{y})$ is the likelihood function for the parameter $\boldsymbol{\theta}$ with respect to the observed data, denoted by the vector or matrix $\boldsymbol{y}$. Bayesian inference employs Bayes' theorem to update the beliefs about the random variable $\boldsymbol{\theta}$ as more information becomes available via the observed data. In the Bayesian paradigm, inference is facilitated via the posterior distribution of $\boldsymbol{\theta}$, denoted by $p(\boldsymbol{\theta}\hspace{1pt}|\hspace{1pt} \boldsymbol{y})$. This distribution communicates which values of $\boldsymbol{\theta}$ are plausible given the observed data and prior beliefs. By Bayes' Theorem, we have that
    \begin{equation}\label{eqn:posterior}
		p(\boldsymbol{\theta}\hspace{1pt}|\hspace{1pt} \boldsymbol{y}) = \dfrac{L(\boldsymbol{\theta};\hspace{1pt} \boldsymbol{y}) \hspace{1pt} p(\boldsymbol{\theta})}{\int L(\boldsymbol{\theta};\hspace{1pt} \boldsymbol{y})\hspace{1pt}p(\boldsymbol{\theta})d\boldsymbol{\theta}} \propto L(\boldsymbol{\theta};\hspace{1pt} \boldsymbol{y})\hspace{1pt}p(\boldsymbol{\theta}).
	\end{equation} 

    To implement fully Bayesian analyses, prior distributions must be specified before observing data. This is in contrast to empirical Bayes methods that set the parameters for the prior distributions to their most likely values given the observed data \citep{casella1985introduction, carlin2000bayes}. In the absence of prior information, uninformative or diffuse priors are often used. When relevant prior information from subject matter experts or previous statistical analyses is available, it rarely takes the form of quantifiable prior distributions on the unobservable parameter(s) of a statistical model. Prior elicitation procedures are used to transfigure prior information into quantifiable prior distributions.

Winkler \citep{winkler1967assessment} conducted some of the initial work on prior elicitation, citing the siloed nature of prior specification and posterior analysis. Most previous Bayesian research had investigated how to leverage sample information from the likelihood function to obtain the posterior from the prior distribution, which was assumed to have been already assessed. Winkler \citep{winkler1967assessment} explored elicitation methods for Bernoulli processes that involved asking questions about the prior cumulative distribution function (CDF) or probability distribution function (PDF) for the Bernoulli parameter. These elicitation methods were for univariate priors, yet Winkler \citep{winkler1967assessment} acknowledged that the assessment of prior distributions in multivariate contexts was an important and nontrivial problem.



Decades later, Chaloner \citep{chaloner1996elicitation} overviewed methods for subjective prior specification with an emphasis on methods for clinical trials. In this context,  Freedman and Spiegelhalter \citep{freedman1983assessment} elicited a prior on the difference in cancer recurrence probability between treatment groups by asking experts about the mode and likely lower and upper bounds for this difference. This prior was not combined with observed data and instead used for design purposes to choose the number of interim analyses in a sequential trial. Other contributions in clinical settings advocated for soliciting prior beliefs from several experts to form a community of prior distributions and basing inference on a consensus of posterior conclusions \citep{kadane1986progress, chaloner1993graphical}. To reduce cognitive and computational complexity, most elicitation methods overviewed by Chaloner \citep{chaloner1996elicitation} relied on parametric assumptions and solicited information about potentially observable conduits for the unobservable parameter $\boldsymbol{\theta}$ that are easier to conceptualize. For instance, priors for Bayesian regression model coefficients were elicited using information about quantiles of predictive distributions \citep{kadane1980interactive} and survival probabilities \citep{chaloner1993graphical}. 




Additional reviews of prior elicitation methods have since been conducted \citep{garthwaite2005statistical, o2006uncertain, johnson2010methods}. In one contribution, Johnson et al. \citep{johnson2010methods} conducted a systematic review of prior elicitation methods with an emphasis on the feasibility of the process in terms of the required time, cost, personnel, and equipment. They found that although expert fatigue and lack of understanding compromise the reliability of elicitation procedures \citep{winkler1971probabilistic, garthwaite2005statistical}, the reviewed methods were rarely formally evaluated on their feasibility. 

Recent contributions have aimed to reduce expert fatigue and improve understanding by developing iterative elicitation procedures that provide experts with instant feedback via a graphical interface \citep{jones2014prior, casement2018graphical, williams2021comparison, casement2023phoropter}. Prior elicitation procedures have also been developed for more complex settings -- including methods for rank analysis \citep{crispino2023informative}, nonparametric models \citep{seo2022nonparametric}, sequential analysis \citep{santos2019algorithm}, power priors \citep{ye2022normalized}, and mixture models \citep{fuquene2019choosing, feroze2021comparison}. One such area of research involves using copulas to accommodate more flexible dependence structures when eliciting multivariate priors \citep{elfadaly2017eliciting, wilson2018specification, wilson2021recent}. These contributions are novel given that the dependence structure is an afterthought in many recent prior elicitation methods. In particular, many recent methods consider elicitation for univariate parameters \citep{casement2018graphical, casement2023phoropter}, fix the dependence structure to leverage conjugacy \citep{santos2019algorithm, srivastava2019subjective}, or assume that the components of $\boldsymbol{\theta}$ are independent a priori \citep{garthwaite2013prior,jones2014prior,fuquene2019choosing,seo2022nonparametric}. Prominent recent advances with copula-based priors are overviewed in Section \ref{sec:int.2}. A primer on copulas is provided in Section \ref{sec:int.cop}.


\subsection{Background on Copula Models}\label{sec:int.cop}

The behaviour of random variables $\boldsymbol{X} = \{X_j\}_{j = 1}^d$ is often characterized by their joint distribution function $H(\boldsymbol{x})$. Each component of $\boldsymbol{X}$ also has a marginal distribution function $F_j(x_j) = Pr(X_j \le x_j), ~ j = 1, \dots, d$. Copulas flexibly allow for the dependence structure of $\boldsymbol{X}$ to be considered separately from its marginals when eliciting multivariate distributions. Let $U_1, U_2, \dots, U_d$ be uniformly-distributed random variables over the unit interval $[0,1]$. The distribution function
	\begin{equation*}\label{eqn:copula}
		C(u_1, \dots, u_d) = Pr(U_1 \le u_1, \dots, U_d \le u_d)
	\end{equation*} 
is such that $C: [0,1]^d \rightarrow [0,1]$ is a copula \citep{nelsen2006introduction}. Sklar's theorem \citep{sklar1959fonctions, schweizer2011probabilistic} explicates the relationship between the copula $C$, the multivariate joint distribution function $H(\boldsymbol{x})$, and the univariate marginal CDFs $F_j(x_j)$ for $j = 1, \dots, d$:
	\begin{equation*}\label{eqn:sklar}
		H(\boldsymbol{x}) = C(F_1(x_1), \dots, F_d(x_d)),
	\end{equation*} 
where $\boldsymbol{x} = (x_1, \dots, x_d) \in \mathcal{X} \subseteq \mathbb{R}^d$. Copulas are therefore incorporated into multivariate distributions even if they are not explicitly defined. If $F_1, \dots, F_d$ are continuous, the copula $C$ is unique.

A copula $C$ can be represented as the sum of its absolutely continuous component $A_C$ and singular component $S_C$ \citep{nelsen2006introduction}. For $\boldsymbol{u} = (u_1, \dots, u_d) \in [0,1]^d$, 
	\begin{equation*}\label{eqn:copula_decomp2}
		A_C(\boldsymbol{u}) = \int_0^{u_1}\cdots \int_0^{u_d}\dfrac{\partial^d}{\partial t_1 \cdots \partial t_d} C(t_1, \dots, t_k) d t_k \cdots d t_1,
	\end{equation*} 
and $S_C(\boldsymbol{u}) = C(\boldsymbol{u}) - A_C(\boldsymbol{u})$. If $C = A_C$ on $[0,1]^d$, the copula $C$ is absolutely continuous and admits a density function 
	\begin{equation*}\label{eqn:copula_den}
		c(\boldsymbol{u}) = \dfrac{\partial^d}{\partial u_1 \cdots \partial u_d} C(u_1, \dots, u_d).
	\end{equation*} 
Moreover, if the support of $C$ is $[0,1]^d$, the copula is deemed to have full support \citep{nelsen2006introduction}. All copulas considered in this paper adhere to this definition.


Copulas can be defined using common probability distributions. For instance, the Gaussian copula \citep{clemen1999correlations} with correlation matrix $\boldsymbol{R}$ is defined such that
	\begin{equation*}\label{eqn:cop_gauss}
		C^{Ga}_{\boldsymbol{R}}(\boldsymbol{u}) = \Phi_{\boldsymbol{R}}(\Phi^{-1}(u_1), \dots, \Phi^{-1}(u_d)),
	\end{equation*} 
where $\Phi$ is the CDF of the $\mathcal{N}(0, 1)$ distribution and $\Phi_{\boldsymbol{R}}: \mathbb{R}^d \rightarrow [0,1]$ is the CDF of the $d$-dimensional $\mathcal{N}(\boldsymbol{0}, \boldsymbol{R})$ distribution with covariance matrix equal to the correlation matrix $\boldsymbol{R}$. The $t$-copula \citep{demarta2005t} can be similarly defined given a multivariate $t$-distribution with correlation matrix $\boldsymbol{R}$ and degrees of freedom $\nu$. In contrast, Archimedean copulas are a commonly used class of copulas that are efficiently parameterized via generator functions \citep{nelsen2006introduction}. This paper focuses on parametric copula models, but copulas can also be leveraged in a nonparametric framework \citep{wong2010optional, wu2015bayesian, ning2018nonparametric, barone2023bayesian}.

\begin{figure}[t] \centering 
		\includegraphics[width = 0.35\textwidth]{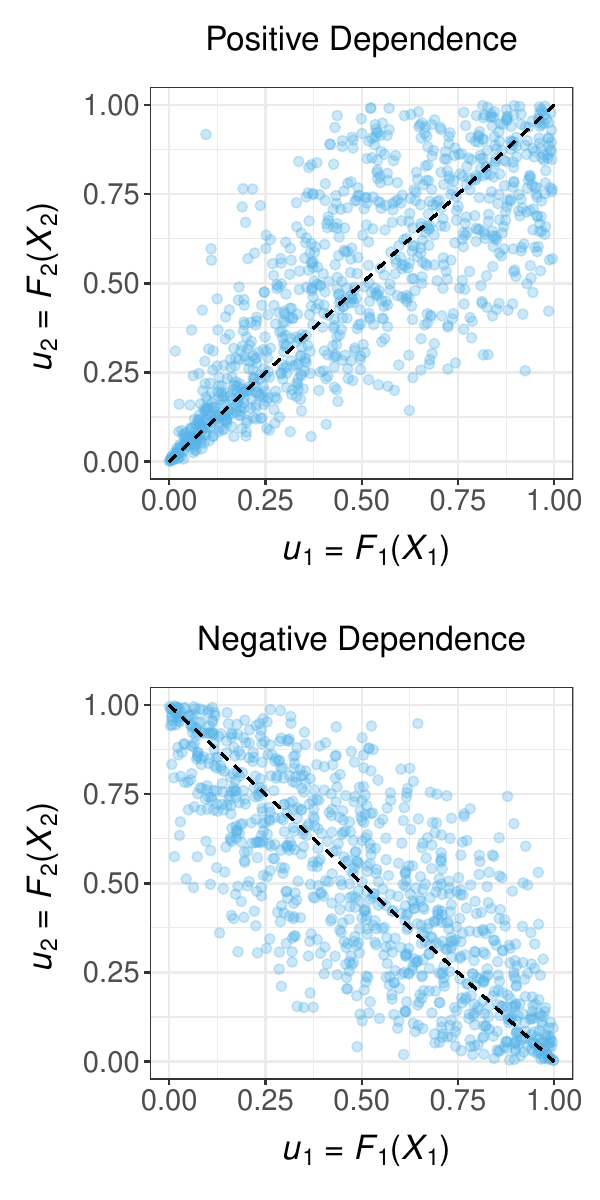} 
		\caption{\label{fig:cop} Samples of 1000 points from a Clayton copula (top) and a Gaussian copula with Pearson's $\rho = -0.8$ (bottom). The upper and lower Fréchet-Hoeffding bounds are given by the dotted lines.} 
	\end{figure}

For $d = 2$ dimensions, Figure \ref{fig:cop} visualizes samples from two copulas. In the top plot, the blue points are generated from an Archimedean Clayton copula with parameter value $\phi = 3 \ge 0$, which characterizes positive dependence. The extent of the dependence between two random quantities is constrained by the Fréchet-Hoeffding bounds \citep{nelsen2006introduction}. The dotted line in the top plot of Figure \ref{fig:cop} is the upper Fréchet-Hoeffding bound. This bound characterizes dependence for comonotonic variables. In the bottom plot, the blue points are generated from a Gaussian copula parameterized by Pearson's $\rho = - 0.8$. The dotted line in the bottom plot is the lower Fréchet-Hoeffding bound. This bound characterizes dependence for countermonotonic variables. As the strength of the positive (negative) dependence between two variables increases, their $(u_1, u_2)$ combinations tend to cluster around the upper (lower) Fréchet-Hoeffding bound. The dependence structure between two independent random variables is characterized by the independence copula: $c(\boldsymbol{u}) = 1$ for $\boldsymbol{u} \in [0,1]^2$. In two dimensions, a sample from this copula approximates random scatter over $[0,1]^2$.

 The upper Fréchet-Hoeffding bound for positive dependence generalizes to settings with more than two dimensions. However, negative dependence in higher dimensions is more complicated since many random variables cannot exhibit strong mutual negative dependence. In addition to characterizing the strength and direction of dependence between random variables, copula models also account for symmetry and tail dependence. For instance, the Clayton copula accounts for lower tail dependence since the points in the top plot of Figure \ref{fig:cop} are tightly clustered in the bottom left corner.

 \subsection{Recent Developments with Copula-Based Priors}\label{sec:int.2}
 
 Copulas can be incorporated into the prior elicitation process as illustrated in recent developments for the multinomial model. Elfadaly and Garthwaite \citep{elfadaly2017eliciting} proposed one such method to elicit Gaussian copula prior distributions. The standard multinomial model assumes that data $y_{i} \in \{1, 2, \dots, w \},~ i = 1,\dots,n$ are collected independently and that the outcome $v$ occurs with probability $0 < p_{v} < 1$ for $v = 1, \dots, w$ such that $\sum_{v = 1}^w p_{v} = 1$. The multinomial model is parameterized by $\boldsymbol{p}_{\color{black}{w}} \color{black}{=} (p_1, \dots, p_w)$. However, Elfadaly and Garthwaite \citep{elfadaly2017eliciting} did not directly assign a Gaussian copula to the multinomial probabilities since that approach would not enforce the unit-sum constraint. Instead, they defined new variables $Z_1, \dots, Z_w$ such that
\begin{equation}\label{eqn:ord_z}
\begin{split}
Z_1 = p_1, ~~~~Z_v = \dfrac{p_v}{1 - \sum_{t=1}^{v-1}p_t}~~\text{for}~~ v = 2, \dots, w-1, \\ \text{and}~~~~ Z_w = 1.
\end{split}
	\end{equation} 
The corresponding inverse transformations are given by
\begin{equation}\label{eqn:ord_p}
		p_1 = Z_1~~~~\text{and}~~~~p_v = Z_v\prod_{t = 1}^{v-1}(1 - Z_t) ~~\text{for}~~v = 2, \dots, w.
	\end{equation} 


\begin{figure*}[!tb] \centering 
		\includegraphics[width = \textwidth]{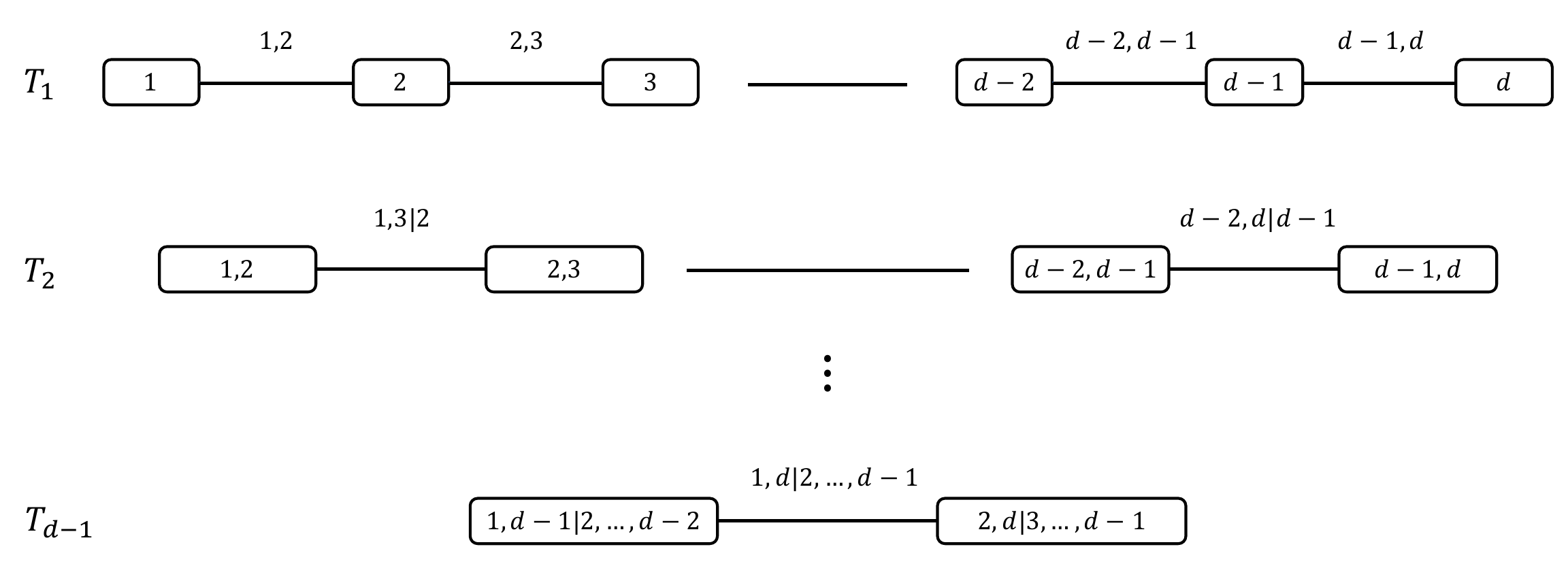} 
		\caption{\label{fig:vine_copula} The structure of a general D-Vine on $d$ variables. \color{black}For each tree $T_j$, the node set $N_j$ inside the rectangles is joined by the edge set $E_j$, labelled using the text over the line segments. These labels take the form $e_1, e_2 \hspace*{2pt}|\hspace*{2pt}D_{\boldsymbol{e}}$.\color{black}} 
	\end{figure*}
 
The variable $Z_v$ represents the probability that an observation is assigned to category $v$ given that it has not been assigned to categories $1, \dots, v-1$. Elfadaly and Garthwaite
\citep{elfadaly2017eliciting} assigned marginal $\text{BETA}(\alpha_v, \beta_v)$ priors to $Z_v$, $v = 1, \dots, w-1$. A joint prior for $\boldsymbol{\theta} = \boldsymbol{Z}_{w-1} = (Z_1, \dots, Z_{w-1})$ that satisfies the unit-sum constraint \color{black}{for $\boldsymbol{p}_w$} \color{black}{was} created by joining the marginal beta priors with a Gaussian copula. If $Z_1, \dots, Z_{w-1}$ are independent, $\boldsymbol{p}_{w}$ follows a  generalized Dirichlet distribution \citep{connor1969concepts}. In this scenario, Gaussian copulas were leveraged to construct priors that were more flexible than standard alternatives.  


Elfadaly and Garthwaite \citep{elfadaly2017eliciting} constructed marginal beta distributions for $Z_v, v = 1, \dots, w-1$ by soliciting estimates for the quartiles of each variable. For each variable, the three quartile estimates were reconciled into a two-parameter beta distribution using least-squares optimization. They formed the correlation matrix $\boldsymbol{R}$ for the Gaussian copula by soliciting further estimates. For $v = 2, \dots, w-1$, experts were asked to update their estimate for the median of $p_v$ under the assumption that the median of $p_{v-1}$ was equal to the lower quartile specified in the previous step. These additional estimates were used in conjunction with a method from Kadane et al. \citep{kadane1980interactive} to ensure $\boldsymbol{R}$ was positive definite. 

If the prior must admit a density function, only absolutely continuous copulas should be considered during the elicitation process. Unless certain combinations of the variables in $\boldsymbol{\theta}$ are invalid, the candidate copulas should have full support so as to not inadvertently restrict the domain of the parameter space $\boldsymbol{\Theta}$ a priori. When $\boldsymbol{R}$ has full rank, the Gaussian copula is absolutely continuous with full support. 


Wilson \citep{wilson2018specification} extended this method for use with vine copulas \citep{joe1996families, bedford2002vines, joe2011dependence}, using a similar process as Elfadaly and Garthwaite \citep{elfadaly2017eliciting} to elicit marginal beta distributions for $Z_v, v = 1, \dots, w-1$. This extended method leveraged D-vines \citep{kurowicka2005distribution} to incorporate more flexibility into the prior dependence structure than the Gaussian copula can accommodate. For a model with $d$ variables, D-vines utilize the graphical structure in Figure \ref{fig:vine_copula} to characterize dependence between the variables in $\boldsymbol{\theta} = (\theta_1, \dots, \theta_d)$ using $d-1$ trees: $T_1, \dots, T_{d-1}$. $T_1$ consists of a node set $N_1 = \{1, 2, \dots, d \}$ and an edge set $E_1 = \{(1,2), (2,3), \dots, (d-1,d) \}$, where the integers in the node and edge sets refer to indices in $\boldsymbol{\theta}$. For $j = 2, 3, \dots, d-1$, the node set of $T_j$ is $N_j = E_{j-1}$, and two edges in $E_{j-1}$ are connected with an edge in $T_j$ only if they share a common node \color{black}{in $N_{j-1}$}\color{black}{.} D-vines characterize dependence in higher-dimensional settings using (un)conditional bivariate copulas. Each edge $\boldsymbol{e}$ in the edge set $E(\mathcal{V}) = \cup_{j = 1}^{d-1}E_j$ considers the dependence between two variables in $\boldsymbol{\theta}$, denoted $e_1$ and $e_2$, conditional on the variables in a potentially empty set $D_{\boldsymbol{e}} \subset \{1,2,\dots, d \}$. \color{black}For an edge $\boldsymbol{e}$, $D_{\boldsymbol{e}}$ consists of all indices that are common to \emph{both} nodes connected by the edge, and the two remaining indices are $e_1$ and $e_2$.


The magnitude and direction of dependence in bivariate copulas is often summarized via Kendall's $\tau$ \citep{kendall1938new} values.  Kendall's $\tau$ measures rank correlation in terms of how similar the orderings of bivariate data are when ranked by each quantity. The D-vine structure allows the bivariate copula corresponding to each edge $\boldsymbol{e} \in E(\mathcal{V})$ to take any Kendall's $\tau$ value that does not correspond to comonotonicity or countermonotonicity \citep{bedford2002vines}: $\tau_{e_1, e_2|D_{\boldsymbol{e}}} \in (-1,1)$. 

When used to specify a prior $p(\boldsymbol{\theta})$, D-vines prompt the set of Kendall's $\tau$ values $\{\tau^p_{e_1, e_2|D_{\boldsymbol{e}}}\}_{\boldsymbol{e} \in E(\mathcal{V})}$. For each bivariate prior copula, its family determines the range of possible values for Kendall's $\tau$. \color{black}Wilson \citep{wilson2018specification} proposed considering Gaussian and $t$-copulas as candidate copulas, along with several Archimedean copulas that are absolutely continuous with full support. These bivariate copulas were selected by soliciting estimates for the conditional quartiles of the $p_v$ and $Z_v$ variables. Across all candidate copula families considered, the best-fitting copula was parameterized to minimize least squares between solicited and induced prior quantiles on the $Z_v$ variables. 


\subsection{Contributions}

This paper provides general guidance for prior dependence specification in multivariate settings. These recommendations are topical given that recent advances in copula-based priors allow for the incorporation of unprecedented flexibility into the prior dependence structure. This additional flexibility could give rise to priors that more accurately characterize real-life phenomena, but we argue that this flexibility is only useful in certain contexts. As such, \color{black}{statisticians should be equipped to help practitioners cull} \color{black}{the} vast number of potential prior dependence structures and select one that aligns with their objectives for posterior analysis. \color{black}{By exploring \emph{how} the prior dependence structure can impact the posterior distribution, our recommendations clarify when it is useful to elicit complicated prior dependence structures and draw inferences regarding posterior dependence.} \color{black}{These} recommendations are illustrated using several models for which copula-based priors have been proposed. Unlike situations that specify a community of prior distributions, we restrict our discussion to the case where a single prior is elicited.


The remainder of this article is structured as follows. In Section \ref{sec:posterior}, we define general conditions under which the prior dependence structure is incompatible with that induced by the likelihood function and unable to be retained by the posterior distribution as data are observed. \color{black}{This result may not surprise Bayesian statisticians. Yet, we argue that it is important to explicate this finding clearly -- both for the sake of practitioners and when teaching students who will eventually guide practitioners as statistics experts. In Section \ref{sec:design}, we emphasize small-sample scenarios in which flexible prior dependence structures can be useful even if they are not retained asymptotically. We also discuss how the inability to retain prior dependence might be used to diagnose philosophical issues with the objectives for posterior analysis.} \color{black}{In} Section \ref{sec:analysis}, we prove asymptotic results about the impact of the prior dependence structure on the convergence of the posterior mode, which we contrast with previous work on copula-based priors \citep{michimae2022bayesian}. We then conduct numerical studies with both small and large sample sizes to contextualize these theoretical results and prompt further recommendations for choosing prior dependence structures. We provide concluding remarks and a discussion of extensions to this work in Section \ref{sec:disc}. 


\section{Retention of Prior Dependence}\label{sec:posterior}

\subsection{Background}\label{sec:post.motivation}

In this section, we examine situations where the prior dependence structure cannot carry over into the posterior distribution as data are collected. It is unrealistic to expect the prior dependence structure for $\boldsymbol{\theta}$ to be perfectly specified. However, we should be mindful of whether the prior dependence structure can be retained a posteriori before using the posterior of $\boldsymbol{\theta}$ to draw conclusions based on its dependence structure. \color{black}{This awareness primes statisticians for questions from practitioners about why we may draw different conclusions given small and large samples generated from the same data generation process. The inability to retain prior dependence might also allude to misalignment between the complexity of the likelihood function and the objectives for posterior analysis.}


\color{black}{The} concept of chronic rejection \citep{libby2001chronic, vos2011randomised} frames this section's main result. The term chronic rejection describes the process in which a transplanted organ is rejected by the recipient's immune system over a long period of time. The recipient's persistent immune response against the transplanted organ causes gradual damage. Similar to screening an organ donor, we argue that one should consider whether the prior dependence structure for $\boldsymbol{\theta}$ is compatible with that induced by the likelihood function. Such incompatibilities imply that the prior dependence structure cannot be retained a posteriori -- and the prior dependence structure is hence a chronically rejected one. We emphasize that the term rejection as used in this paper does not imply the formal rejection of a statistical hypothesis test. In Section \ref{sec:chronic}, we define the notion of chronically rejected prior dependence structures. We illustrate this definition \color{black}{with priors elicited using methods from Elfadaly and Garthwaite \citep{elfadaly2017eliciting} and Wilson \citep{wilson2018specification}} \color{black}{in} Section \ref{post.illustration}. \color{black}{To} help inform conversations between statisticians and practitioners, Section \ref{sec:design} discusses whether the use of chronically rejected prior dependence structures aligns with various posterior objectives.\color{black}{}



\subsection{Chronically Rejected Prior Dependence Structures}\label{sec:chronic}


Here, we define general conditions under which prior dependence structures cannot be retained by the posterior with enough data. These sufficient conditions can be readily verified prior to observing data \color{black}{to} assess the alignment of the prior dependence structure with posterior objectives. \color{black}{We} consider the limiting behaviour of the posterior distribution for $\boldsymbol{\theta}$. Because we consider this behaviour under various data generation processes, the data are random variables. Data from a random sample of size $n$ are represented by $\boldsymbol{Y}^{_{(n)}}$. Realizations of these samples are denoted by $\boldsymbol{y}^{_{(n)}}$. Much of the work on limiting posteriors  (see e.g., \citep{ghosal1995convergence}, \citep{le2000asymptotics}, \citep{gao2020general}, or \citep{schillings2020convergence}) appeals to the Bernstein-von Mises theorem \citep{vaart1998bvm}. We let $m(\hspace{1pt} \cdot \hspace{1pt}| \hspace{1pt}\boldsymbol{\theta})$ be the statistical model corresponding to the likelihood function in (\ref{eqn:posterior}). When data $\boldsymbol{Y}^{_{(n)}}$ are generated independently and identically from $m(\hspace{1pt} \cdot \hspace{1pt}| \hspace{1pt}\boldsymbol{\theta}_0)$, the Bernstein-von Mises theorem dictates that the posterior for $\boldsymbol{\theta}$ converges to the $\mathcal{N}(\boldsymbol{\theta}_0, \mathcal{I}(\boldsymbol{\theta}_0)^{-1}/n)$ distribution in the limit of infinite data, where $\mathcal{I}(\cdot)$ is the Fisher information. 

In addition to the independently and identically distributed assumption, there are three conditions that must be satisfied to invoke the Bernstein-von Mises theorem. The first two conditions involve the model $m(\hspace{1pt} \cdot \hspace{1pt}| \hspace{1pt}\boldsymbol{\theta}_0)$, and they are collectively weaker than the conditions for the asymptotic normality of the maximum likelihood estimator (MLE) \citep{lehmann1998theory}. Condition 1 ensures the model $m(\hspace{1pt} \cdot \hspace{1pt}| \hspace{1pt}\boldsymbol{\theta}_0)$ is differentiable in quadratic mean with nonsingular $\mathcal{I}(\boldsymbol{\theta}_0)$. Condition 2 requires that there exist a sequence of uniformly consistent tests for $H_0: \boldsymbol{\theta} = \boldsymbol{\theta}_0$ vs. $H_1: \lVert \boldsymbol{\theta} - \boldsymbol{\theta}_0 \rVert_2 \ge \varepsilon$ for every $\varepsilon > 0$. Condition 3 concerns the prior distribution $p(\boldsymbol{\theta})$ used to analyze the observed data. This prior must be absolutely continuous in a neighbourhood of $\boldsymbol{\theta}_0$ with $p(\boldsymbol{\theta}_0) > 0$. We consider priors defined such that 
\begin{equation}\label{eqn:prior}
p(\boldsymbol{\theta}) = \prod_{j = 1}^{d}f_j(\theta_j) \times c(F_1(\theta_1), \dots, F_{d}(\theta_{d})),
	\end{equation} 
where $c(u_1, \dots, u_{d})$ is the copula density function for an absolutely continuous copula $C$ and $F_j(\theta_j)$ and $f_j(\theta_j)$ are respectively the marginal prior CDF and PDF for $\theta_j$, $j = 1, \dots, d$. 

To directly apply the Bernstein-von Mises theorem, a single value of $\boldsymbol{\theta}_0 \in \boldsymbol{\Theta}$ must be selected. It is not realistic to expect practitioners to correctly identify this value for $\boldsymbol{\theta}_0$ a priori. As such, our results incorporate uncertainty about the value of $\boldsymbol{\theta}_0$ used to generate $\boldsymbol{Y}^{_{(n)}}$. We do so by introducing a prior $p_D(\boldsymbol{\theta})$ that defines the prior predictive distribution of $\boldsymbol{Y}^{_{(n)}}$: \begin{equation}\label{eq:prior_pred}
p(\boldsymbol{y}^{_{(n)}}) = \int \prod_{i}^nm(y_i \hspace{1pt}| \hspace{1pt} \boldsymbol{\theta})\hspace{1pt} p_D(\boldsymbol{\theta}) \hspace{1pt} d\boldsymbol{\theta}.
\end{equation}
 
In (\ref{eq:prior_pred}), data $\boldsymbol{Y}^{_{(n)}}$ are generated independently and identically from $m(\hspace{1pt} \cdot \hspace{1pt}| \hspace{1pt}\boldsymbol{\theta}_0)$ given $\boldsymbol{\theta}_0 \sim p_D(\boldsymbol{\theta})$. This data generation process allows us to compare various objectives for prior specification using repeated simulation in Section \ref{sec:design}. The prior in (\ref{eq:prior_pred}) may or may not be the same prior as $p(\boldsymbol{\theta})$ used to analyze the observed data in (\ref{eqn:prior}), which is often called the analysis prior. To explore the limiting behaviour of the posterior when data are generated via (\ref{eq:prior_pred}), we note that the Bernstein-von Mises theorem considers a special case of $p(\boldsymbol{y}^{_{(n)}})$ in which the prior $p_D(\boldsymbol{\theta})$ is degenerate. That is, $p_D(\boldsymbol{\theta}_0) = 1$ for a particular $\boldsymbol{\theta}_0 \in \boldsymbol{\Theta}$, and 0 otherwise. In light of this, we emphasize that the prior that must satisfy condition 3 for the Bernstein-von Mises theorem is the analysis prior $p(\boldsymbol{\theta})$. Theorem \ref{thm1} generalizes results from the Bernstein-von Mises theorem to nondegenerate priors $p_D(\boldsymbol{\theta})$ under certain conditions.


 \begin{theorem}\label{thm1}
Let $\boldsymbol{\Theta}^*$ be the set of interior points in $\boldsymbol{\Theta}$. Suppose conditions 1, 2, and 3 for the Bernstein-von Mises theorem are satisfied for all $\boldsymbol{\theta} \in \boldsymbol{\Theta}^*$. Let data $\boldsymbol{Y}^{_{(n)}}$ be generated via (\ref{eq:prior_pred}) such that $\boldsymbol{\theta}_0 \sim p_D(\boldsymbol{\theta})$ and $p_D(\boldsymbol{\theta}) = 0$ for all $\boldsymbol{\theta} \notin \boldsymbol{\Theta}^*$. The posterior dependence structure of $\boldsymbol{\theta}$ given observed $\boldsymbol{y}^{_{(n)}}$ converges to $C^{Ga}_{\boldsymbol{R}}(\cdot)$ corresponding to the covariance matrix $\mathcal{I}(\boldsymbol{\theta}_0)^{-1}$ as $n \rightarrow \infty$.
\end{theorem}


\textbf{Proof of Theorem 1}. When the conditions for Theorem \ref{thm1} hold, $p(\boldsymbol{\theta}\hspace{1pt}|\hspace{1pt} \boldsymbol{y}^{_{(n)}})$ is approximately multivariate normal with covariance matrix $\mathcal{I}(\boldsymbol{\theta}_0)^{-1}/n$ for large enough $n$. The posterior dependence structure of $\boldsymbol{\theta}$ is hence reasonably characterized by a Gaussian copula with correlation matrix $\boldsymbol{R}$ corresponding to $\mathcal{I}(\boldsymbol{\theta}_0)^{-1}$. This result follows by the Bernstein-von Mises theorem because we restrict the value of $\boldsymbol{\theta}_0$ to be contained in $\boldsymbol{\Theta}^*$; the Bernstein-von Mises theorem is applicable when $\boldsymbol{\theta}_0$ is not a boundary point of the parameter space $\boldsymbol{\Theta}$.$\hspace*{-5pt}\qed$


Under the conditions in Theorem \ref{thm1}, \color{black}{the posterior copula for $\boldsymbol{\theta}$ generally becomes more limited in its flexibility to characterize dependence. To see this, notice that} \color{black}{the} posterior copula for $\boldsymbol{\theta}$ is approximately Gaussian for large samples. \color{black}{Moreover, the magnitude and direction of the dependence in this approximately Gaussian copula is constrained by the likelihood function through the matrix $\mathcal{I}(\boldsymbol{\theta}_0)^{-1}$ and possible values for $\boldsymbol{\theta}_0 \in \boldsymbol{\Theta}^*$.} \color{black}{Yet,} we may not collect nearly enough data for this copula to be Gaussian in practice. 

Because it is unrealistic to expect the prior dependence structure of $\boldsymbol{\theta}$ to be perfectly specified, we consider a partial characterization of the dependence structure \color{black}{based on Kendall's $\tau$ values for} \color{black}{D-vines} in Corollary \ref{cor2}. Even if the copula in (\ref{eqn:prior}) is not specified using a D-vine, such a structure can be induced. \color{black}{We} let $\{\tau^p_{e_1, e_2|D_{\boldsymbol{e}}}\}_{\boldsymbol{e} \in E(\mathcal{V})}$ defined in Section \ref{sec:int.2} characterize the prior dependence structure on $\boldsymbol{\theta}$. \color{black}{The matrix $\mathcal{I}(\boldsymbol{\theta}_0)^{-1}$ and standard results about conditional normality prompt limiting Kendall's $\tau$ values for each pair of variables $e_1$ and $e_2$ in $\boldsymbol{\theta}$ given the variables in their conditioning set $D_e$: $\{\tau(\boldsymbol{\theta}_0)_{e_1, e_2|D_{\boldsymbol{e}}}\}_{\boldsymbol{e} \in E(\mathcal{V})}$.}

\color{black}

\begin{corollary}\label{cor2}
    Under the conditions for Theorem \ref{thm1}, let $\{\tau^p_{e_1, e_2|D_{\boldsymbol{e}}}\}_{\boldsymbol{e} \in E(\mathcal{V})}$ describe prior dependence on $\boldsymbol{\theta}$. Suppose no $\boldsymbol{\theta}_0$ with $p_D(\boldsymbol{\theta}_0) > 0$ is such that $C^{Ga}_{\boldsymbol{R}}(\cdot)$ corresponding to the covariance matrix $\mathcal{I}(\boldsymbol{\theta}_0)^{-1}$ induces a dependence structure $\{\tau(\boldsymbol{\theta}_0)_{e_1, e_2|D_{\boldsymbol{e}}}\}_{\boldsymbol{e} \in E(\mathcal{V})}$ such that $\tau(\boldsymbol{\theta}_0)_{e_1, e_2|D_{\boldsymbol{e}}} = \tau^p_{e_1, e_2|D_{\boldsymbol{e}}}$ for all $\boldsymbol{e} \in E(\mathcal{V})$. Then, the posterior of $\boldsymbol{\theta} \hspace{1pt}|\hspace{1pt} \boldsymbol{Y}^{_{(n)}}$ cannot retain the magnitude and direction of prior dependence as $n \rightarrow \infty$.
\end{corollary}


Corollary \ref{cor2} follows directly from Theorem \ref{thm1}. We suppose there is no $\boldsymbol{\theta}_0$ with $p_D(\boldsymbol{\theta}_0) > 0$ such that the Gaussian copula corresponding to the covariance matrix $\mathcal{I}(\boldsymbol{\theta}_0)^{-1}$ induces a dependence structure characterized by $\{\tau^p_{e_1, e_2|D_{\boldsymbol{e}}}\}_{\boldsymbol{e} \in E(\mathcal{V})}$. It follows that the magnitude and direction of the dependence structure for $p(\boldsymbol{\theta}\hspace{1pt}|\hspace{1pt} \boldsymbol{y}^{_{(n)}})$ cannot be characterized by $\{\tau^p_{e_1, e_2|D_{\boldsymbol{e}}}\}_{\boldsymbol{e} \in E(\mathcal{V})}$ for sufficiently large $n$. We emphasize that considering dependence structures via Kendall's $\tau$ on the D-vine structure of the copula does \emph{not} fully specify the dependence structure. This allows for more flexibility in the choices for the copula families; it also facilitates the consideration of prior dependence for subvectors of $\boldsymbol{\theta}$, which is useful because it may require too much cognitive complexity to assess the full prior dependence structure. However, two bivariate copulas may have the same Kendall's $\tau$ but different properties in terms of symmetry and tail dependence. We focus on the magnitude and direction of dependence to present generally applicable guidance for prior specification \color{black}{in Section \ref{sec:design}}\color{black}.   

Corollary \ref{cor2} \color{black}{facilitates} the straightforward identification of \color{black}{dependence} structures for $\boldsymbol{\theta}$ that cannot be retained as data $\boldsymbol{y}^{_{(n)}}$ are observed. \color{black}{This} identification can diagnose potential disagreement between the degree of flexibility in the prior dependence structure and that induced by the likelihood function. \color{black}{We} refer to prior dependence structures that satisfy the conditions for Corollary \ref{cor2} as chronically rejected dependence structures. The conditions for Corollary \ref{cor2} are sufficient in that there is \emph{no} value for $\boldsymbol{\theta}_0$ with $p_D(\boldsymbol{\theta}_0) > 0$ such that these dependence structures will be retained as data are generated via (\ref{eq:prior_pred}). However, it is not guaranteed that the prior dependence structure will be retained when the conditions for Corollary \ref{cor2} are not satisfied. That is, the true value of $\boldsymbol{\theta}_0$ might not be one of those that prevent the conditions for Corollary \ref{cor2} from being satisfied. Thus, a prior dependence structure that is not retained is not necessarily a chronically rejected one. In this paper, the notion of chronic rejection defines a specific class of prior dependence structures, whereas retention of the prior dependence structure is viewed more generally as a posterior outcome.

\subsection{Illustration with Copula-Based Priors for the Multinomial Model}\label{post.illustration}

We now apply Corollary \ref{cor2} with the multinomial model. In Appendix A of the supplementary materials, we show that the inverse Fisher information matrix for the multinomial model parameterized in terms of $\boldsymbol{\theta} = (Z_1, Z_2, \dots, Z_{w-1})$ from (\ref{eqn:ord_z}) is diagonal for all possible $(Z_1, Z_2, \dots, Z_{w-1}) \in \boldsymbol{\Theta}^* = (0,1)^{w-1}$. \color{black}{That} is, $\tau(\boldsymbol{\theta}_0)_{e_1, e_2|D_{\boldsymbol{e}}} = 0$ for all $\boldsymbol{e} \in E(\mathcal{V})$ and $\boldsymbol{\theta}_0 \in \boldsymbol{\Theta}^*$. For distributions in the exponential family \citep{lehmann1998theory}, conditions 1, 2, and 3 for the Bernstein-von Mises theorem are satisfied as long as the prior $p(\boldsymbol{\theta})$ has support over $\boldsymbol{\Theta}^*$. If $p(\boldsymbol{\theta})$ incorporates any positive or negative dependence between the conditional multinomial probabilities, then $ \tau^p_{e_1, e_2|D_{\boldsymbol{e}}} \ne 0$ for some $\boldsymbol{e} \in E(\mathcal{V})$. By Corollary \ref{cor2}, all such dependence structures are therefore chronically rejected ones. 

\color{black}{For} $w = 3$ categories, we illustrate this phenomenon when combining such priors with the standard multinomial likelihood, which is parameterized by conditional probabilities $\boldsymbol{\theta} = (Z_1, Z_2)$. We specify $p(\boldsymbol{\theta})$ as in (\ref{eqn:prior}) where the marginal prior for $Z_1$ is $\text{BETA}(20,40)$, the marginal prior for $Z_2$ is $\text{BETA}(30,30)$, and these marginal priors are joined using a Gaussian copula parameterized with Pearson's $\rho = -0.9$. This prior distribution conveys that we expect the multinomial probabilities to be roughly equal for all three categories. This prior specification could be facilitated using either of the methods by Elfadaly and Garthwaite \citep{elfadaly2017eliciting} or Wilson \citep{wilson2018specification}. We note that the prior copula gives rise to a value for Kendall's $\tau$ of $2\text{sin}^{-1}(-0.9)/\pi = -0.713$.

For each of 10000 simulation repetitions, we generated $\boldsymbol{\theta}_0 = (Z_{1,0}, Z_{2,0})$ from the prior specified in the previous paragraph. We generated $\boldsymbol{Y}^{_{(n)}}$ for $n = 10$ from the multinomial model parameterized by $\boldsymbol{\theta}_0$. For each sample, we approximated the posterior of $\boldsymbol{\theta} \hspace{1pt} | \hspace{1pt} \boldsymbol{y}^{_{(n)}}$ using sampling-resampling methods \citep{rubin1988using}, where the proposal distribution was the posterior of $\boldsymbol{\theta} \hspace{1pt} | \hspace{1pt} \boldsymbol{y}^{_{(n)}}$ obtained by independently joining the marginal priors for $Z_1$ and $Z_2$. For each posterior sample, we estimated Kendall's $\tau$ for $Z_1$ and $Z_2$. This process was repeated for $n = \{10^2, 10^3, 10^4, 10^5\}$. The range of Kendall's $\tau$ values observed in this numerical study is summarized in Table \ref{tab:tau}.


This table shows that the posterior cannot retain the negative dependence structure present in the prior as data are observed from the prior predictive distribution. Even though the analysis prior $p(\boldsymbol{\theta})$ coincides with that used to define the prior predictive distribution of $\boldsymbol{Y}^{_{(n)}}$, the multinomial likelihood function cannot accommodate this dependence between $Z_1$ and $Z_2$. \color{black}{Corollary} \ref{cor2} identifies this disagreement between the level of flexibility in the prior dependence structure and that induced by the simple multinomial likelihood function \emph{before} observing data. \color{black}{}

\begin{table}
\centering
\caption{Kendall's $\tau$ values for 10000 posteriors of $Z_1$ and $Z_2$ across various sample sizes $n$}
\label{tab:tau}
\begin{tabular}{@{}lccc@{}}
\hline
& \multicolumn{3}{c}{Kendall's $\tau$} \\[1pt]
\cline{2-4} \\[-6.1pt]
$n$ & \multicolumn{1}{c}{Minimum}
& \multicolumn{1}{c}{Median} & \multicolumn{1}{c}{Maximum} \\
\hline
$10^1$ & -0.7128 & -0.6806 & -0.6638 \\
        $10^2$ & -0.5488 & -0.5049 & -0.4815 \\
        $10^3$ & -0.2103 & -0.1611 & -0.1288 \\
        $10^4$ & -0.0464 & -0.0214 & 0.0062 \\
        $10^5$ & -0.0326 & -0.0022 & 0.0247 \\ \hline
\end{tabular}
\end{table}



The copula-based priors proposed by Elfadaly and Garthwaite \citep{elfadaly2017eliciting} and Wilson \citep{wilson2018specification} define valid posteriors for $\boldsymbol{\theta}$ when combined with the multinomial likelihood, but Table \ref{tab:tau} corroborates that we would draw vastly different conclusions about the posterior dependence structure of $\boldsymbol{\theta}$ for small and large samples. \color{black}{While} Bayesian statisticians may anticipate such results, they may be unexpected by practitioners whom these results materially impact when deciding between prior dependence structures and how much effort to invest in the elicitation process. \color{black}{Elfadaly} and Garthwaite \citep{elfadaly2017eliciting} provided R code to combine their priors with the multinomial likelihood function, and Wilson et al. \citep{wilson2021recent} stated that these priors could be used for Bayesian analysis. However, neither contribution clearly disclosed that these more flexible prior dependence structures cannot be retained as multinomial data are collected. \color{black}{This} example segues into a broader discussion about when the inability to retain the prior dependence structure presents or reveals practical issues with posterior analysis in Section \ref{sec:design}. \color{black}{}

\section{Objectives for Dependence Structure Specification}\label{sec:design}

\color{black}{\subsection{Overview}\label{sec:design.int}}

\color{black}

This section outlines several objectives that practitioners may wish to achieve when specifying prior distributions and their dependence structures. \color{black}{Our} discussion is motivated by the following questions. When is it useful to encode prior dependence that is at odds with the asymptotic covariance structure? When is it sensible to draw inferences about the posterior dependence structure of $\boldsymbol{\theta}$? How are posterior inferences impacted by the dependence structure? Does the inability to retain prior dependence reveal philosophical issues with the objectives for posterior analysis? 

The relevance of each question depends on the objective for prior dependence specification, but we consider retention of the prior dependence structure as defined in Corollary \ref{cor2} for all objectives. The objectives explored in this section are not necessarily mutually exclusive. Nevertheless, we discuss why it may not be possible for a given prior dependence structure to facilitate certain combinations of objectives for posterior analysis. This discussion can guide conversations between statisticians and practitioners to simplify the prior specification process where possible and promote alignment between the prior dependence structure and posterior objectives.\color{black}

\subsection{Supplementation of Small Samples}\label{sec:design.small}

Even if a prior dependence structure cannot be retained as the sample size $n$ increases, it may be used to supplement information from small samples. The following example that illustrates the utility of chronically rejected dependence structures for small samples was helpfully provided by a reviewer of this paper. We suppose that a practitioner aims to model the heights of a new animal species with a $\mathcal{N}(\mu, \sigma^2)$ distribution. The practitioner does not know the typical heights for members of this species, but it is reasonable to expect that $\mu$ and $\sigma$ are on the same scale. Hence, positive prior dependence between $\mu$ and $\sigma^2$ is sensible. Upon observing a single member of this species, the practitioner can estimate $\mu$ -- even though the single observation does not directly provide information about $\sigma^2$. If $\mu$ and $\sigma^2$ were positively correlated a priori, the posterior could convey that a value of one nanometer for $\sigma$ would be unlikely given an observed height of one meter.


For this model, the inverse Fisher information matrix $\mathcal{I}(\boldsymbol{\theta}_0)^{-1}$ is diagonal for all possible values of $\boldsymbol{\theta}_0 = (\mu_0, \sigma^2_0) = \mathbb{R} \times \mathbb{R}^+$. \color{black}{That} is, $\tau(\boldsymbol{\theta}_0)_{\mu, \sigma^2} = 0$ for all $\boldsymbol{\theta}_0 \in \boldsymbol{\Theta}*$. \color{black}{It} follows by Corollary \ref{cor2} that the joint posterior of $\mu$ and $\sigma^2$ cannot retain a prior dependence structure for which Kendall's $\tau$ is not 0. Under the conditions for Corollary \ref{cor2}, any attempt to inject positive or negative dependence into this posterior will be unsuccessful in the limit of infinite data. Once both $\mu$ and $\sigma^2$ are precisely estimated from the data, there will be no practical correlation between their estimates. However, imposing prior dependence is nevertheless useful for small sample sizes.


\color{black}{When chronically rejected dependence structures are used to supplement information from small samples,} we argue that drawing inferences about posterior dependence is problematic. These inferences would just confirm our beliefs about the prior dependence structure despite its incompatibility with likelihood function -- and therefore the data. Such priors are specified for objectives that supersede learning about the dependence structure. For this example, the prior dependence structure was selected with the hope of obtaining a better marginal posterior for $\sigma^2$ with limited information. \color{black}{This} example also suggests that it is not worthwhile to spend time eliciting complicated prior dependence structures that cannot be retained when collecting large samples. Discarding chronically rejected dependence structures can simplify the prior specification process for practitioners in large-sample settings. 

\color{black}{The} impact of posterior dependence on inference is of course not limited to conclusions about the dependence structure of $\boldsymbol{\theta}$. For instance, the dependence structure impacts the posterior of a function of the parameters in $\boldsymbol{\theta}$ and posterior predictive inference. In this example, the posterior of the 0.9 quantile of the $\mathcal{N}(\mu, \sigma^2)$ distribution would be a function of both components in $\boldsymbol{\theta}$. This posterior incorporates information from the posterior dependence structure, but it also critically relies on the marginal posteriors of $\boldsymbol{\theta}$ that we prioritized over the dependence structure via prior specification. As such, we believe that these inferences are more philosophically consistent with our objectives for prior specification than those based exclusively on the dependence structure of $\boldsymbol{\theta}$. \color{black}


\subsection{Coverage of Credible Sets}\label{sec:design.cov}

Prior dependence structures may also be specified to calibrate Bayesian credible sets. In this context, Gustafson \citep{gustafson2012behaviour} referred to the data generating prior as ``nature's'' prior. The data generation prior in our framework is $p_D(\boldsymbol{\theta})$ from (\ref{eq:prior_pred}). If the analysis prior $p(\boldsymbol{\theta})$ coincides with ``nature's'' prior and the model $m(\hspace{1pt} \cdot \hspace{1pt}| \hspace{1pt}\boldsymbol{\theta})$ is correctly specified, credible sets attain their nominal coverage  \citep{gustafson2012behaviour}. We let $\mathcal{J}_{\boldsymbol{\theta}, 1-\alpha}(\boldsymbol{y}^{_{(n)}})$ denote a credible set for $\boldsymbol{\theta}$ given the observed data $\boldsymbol{y}^{_{(n)}}$ with \color{black}{posterior probability of} \color{black}{$1 - \alpha$}. A popular choice for the credible set is that of highest posterior density (HPD). Credible sets attain their nominal coverage when
\begin{equation}\label{eq:nom_cov}
Pr(\boldsymbol{\theta_0} \in \mathcal{J}_{\boldsymbol{\theta}, 1-\alpha}(\boldsymbol{Y}^{_{(n)}})) = 1 - \alpha,
\end{equation}
where $\boldsymbol{Y}^{_{(n)}} \sim m(\hspace{1pt} \cdot \hspace{1pt}| \hspace{1pt}\boldsymbol{\theta}_0)$ such that $\boldsymbol{\theta}_0$ is drawn from $p_D(\boldsymbol{\theta})$. The probabilistic statement in (\ref{eq:nom_cov}) is therefore made with respect to repeated sampling from the prior predictive distribution of $\boldsymbol{Y}^{_{(n)}}$. Once $\boldsymbol{y}^{_{(n)}}$ is observed, the credible set is not random and either contains or does not contain a given value $\boldsymbol{\theta_0}$.

\begin{figure}[tb] \centering 
		\includegraphics[width = 0.475\textwidth]{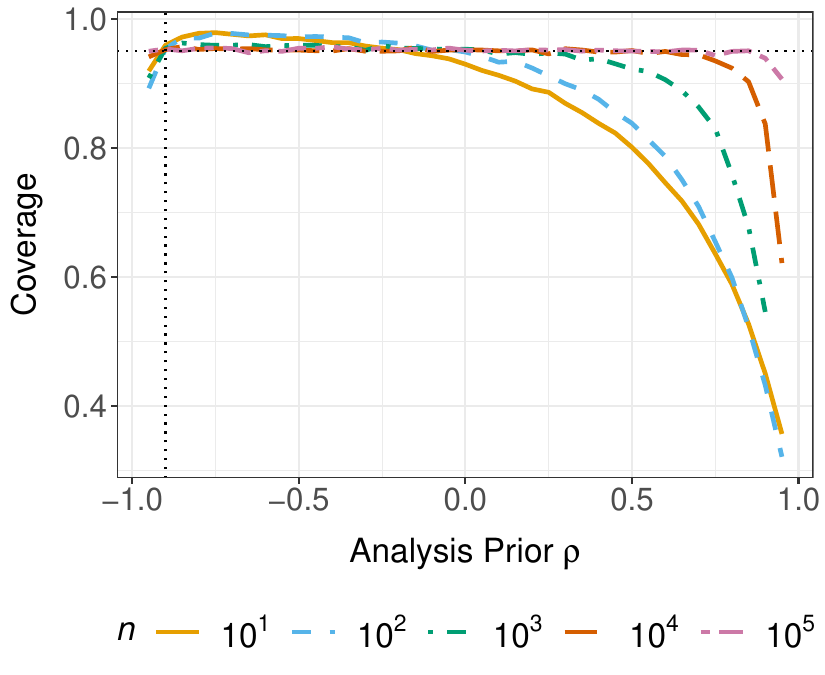} 
		\caption{\label{fig:cov} Empirical coverage of 95\% HPD sets for the multinomial parameter $\boldsymbol{\theta} = (Z_1, Z_2)$ across 10000 posteriors. The horizontal dotted line denotes the nominal coverage, and the vertical one denotes ``nature's'' prior.} 
	\end{figure}

We now investigate how the prior dependence structure impacts \color{black}inference via \color{black}the calibration of credible sets for the multinomial example from Section \ref{post.illustration}. We used the same prior predictive distribution of $\boldsymbol{Y}^{_{(n)}}$ for this numerical study, joining the $\text{BETA}(20,40)$ prior for $Z_1$ and $\text{BETA}(30,30)$ prior for $Z_2$ with a Gaussian copula parameterized with Pearson's $\rho = -0.9$. For each of 10000 simulation repetitions, we approximated the posterior of $\boldsymbol{\theta} \hspace{1pt} | \hspace{1pt} \boldsymbol{y}^{_{(n)}}$ as described in Section \ref{post.illustration} with ``nature's'' analysis prior. For each posterior, we approximated its 95\% HPD set for $Z_1$ and $Z_2$ using two-dimensional kernel density estimation \citep{ripley2002modern}. Empirical coverage was estimated as the proportion of simulation repetitions for which the parameter value $\boldsymbol{\theta}_0 = (Z_{1,0}, Z_{2,0})$ used to generate the multinomial data was contained in this HPD set. We implemented this process for $n = \{10^1, 10^2, 10^3, 10^4, 10^5\}$. We then repeated this process for analysis priors $p(\boldsymbol{\theta})$ that joined the marginal beta priors from ``nature's'' prior with a Gaussian copula parameterized by Pearson's $\rho = \{-0.95, -0.85, -0.80, \dots, 0.95\}$. The results from this numerical study are visualized in Figure \ref{fig:cov}.

 The fact that all five colored curves in Figure \ref{fig:cov} roughly intersect the horizontal dotted line when $\rho = - 0.9$ confirms the result in (\ref{eq:nom_cov}). For small sample sizes $n$, empirical coverage generally deviated from the nominal coverage when $\rho \ne - 0.9$. These deviations resulted in worse coverage when $p(\boldsymbol{\theta})$ was specified such that the magnitude of the negative dependence between $Z_1$ and $Z_2$ was overstated or when the direction of the dependence was inaccurate. The empirical coverage was greater than the nominal coverage when $p(\boldsymbol{\theta})$ incorporated slightly weaker negative dependence between $Z_1$ and $Z_2$ than ``nature's'' prior. For those settings, the copula density function for $p(\boldsymbol{\theta})$ is flatter on $[0,1]^2$ than the copula density function that defines ``nature's'' prior. The resulting credible sets cover a greater range of $\boldsymbol{\theta}$ values but still account for the correct direction of dependence between $Z_1$ and $Z_2$.

 
 As the sample size $n$ increases, the impact of prior dependence is reduced and the empirical coverage approaches the nominal coverage for all $\rho$ values considered. The insights drawn from Figure \ref{fig:cov} are not solely applicable to chronically rejected prior dependence structures. We observed similar results for a numerical study with the gamma model involving analysis priors that do not satisfy the conditions for chronic rejection in Corollary \ref{cor2}. These results are detailed in Appendix B of the online supplement.

 \color{black}{If} a chronically rejected prior dependence structure is used with the intent of calibrating credible sets, we do not advise drawing inferences based solely on the posterior dependence structure. Prior dependence has been specified to prioritize the objective of calibration over the goal of learning about the dependence structure. Moreover, \color{black}{we} suggest erring on the side of understating the strength of dependence structures a priori. This recommendation is not made to encourage practitioners to engineer credible sets with coverages that exceed their nominal values. This excess coverage may not be desirable and cannot persist as the sample size increases. We instead make this recommendation to mitigate the potential ramifications of overstating the strength of dependence structures on the calibration of credible sets. The notion of harmful priors has been discussed elsewhere (see e.g., \citep{reimherr2021prior}), but priors that are harmful to the calibration of credible sets are not necessarily harmful priors in alternative contexts.

\subsection{Inference Regarding Dependence Structures}




\color{black}{We} have argued that drawing inference regarding posterior dependence is problematic when chronically rejected prior dependence structures are specified to achieve a superseding objective. However, drawing inference about the dependence structure may be our main posterior objective or equally as important as other objectives. \color{black}{In} that case, we recommend using Corollary \ref{cor2} to determine whether the prior dependence structure can be retained. \color{black}{It would be ideal if Bayesian inference informed by small and large samples could feasibly support our prior notions about the dependence structure. If the data via the likelihood function cannot possibly support these notions, we should ask practitioners whether the likelihood function or prior distribution require modification.}


\color{black}{Corollary} \ref{cor2} could therefore help diagnose philosophical issues with posterior objectives. While copula-based priors may be able to accommodate flexible dependence structures, the statistical models chosen for the likelihood function may not have this capability. For the multinomial model, there is a single component for each observation of $\boldsymbol{y}^{_{(n)}} = \{y_i \}_{i = 1}^n$. It would be difficult to specify a likelihood function that could accommodate complex dependence structures between the conditional multinomial probabilities given the available data. Thus, it is not always sensible to make inferences about dependence structures. 

\color{black}{This} lack of flexibility alludes to the trade-offs of using simple likelihoods instead of more realistic ones. For instance, Wilson \citep{wilson2018specification} considered the capacity of an engineering structure. These outcomes are influenced by several complex processes that include structural aging and the procedures to assess structure capacity. By representing these outcomes as multinomial, the complexity of the data generating process is simplified, which limits the scope of the conclusions that can be drawn. Gelman et al. \citep{gelman2017prior} argued that the prior can only be interpreted in the context of the likelihood function with which it is paired. This example also underscores the benefits of developing the prior and likelihood function in tandem.  



\color{black}{If} each observation of $\boldsymbol{y}^{_{(n)}}$ is comprised of multiple components, incorporating a copula into the likelihood function may promote greater coherence between the prior and posterior dependence structures. For illustration, we suppose that $\boldsymbol{y}^{_{(n)}} = \{(y_i, y^*_i) \}_{i = 1}^n$, where $y_i \sim \text{EXP}(\lambda)$, $y^*_i \sim \text{EXP}(\kappa)$, and $\lambda$ and $\kappa$ are rates. The likelihood function for this example is such that
\begin{equation}\label{eq:cop_lik}
\begin{split}
 L(\boldsymbol{\theta}; \hspace{1pt} \boldsymbol{y}^{_{(n)}})  \propto \prod_{i = 1}^n & \lambda e^{-\lambda y_i} \times \kappa e^{-\kappa y^*_i} \times \\ & c(1-e^{-\lambda y_i}, 1-e^{-\kappa y^*_i}; \boldsymbol{\upsilon}),
\end{split}
\end{equation}
where the copula density function is parameterized by $\boldsymbol{\upsilon}$ and $\boldsymbol{\theta} = (\lambda, \kappa, \boldsymbol{\upsilon})$. If $c(u_1, u_2; \boldsymbol{\upsilon})$ corresponds to the independence copula, the posterior correlation between $\lambda$ and $\kappa$ based on the inverse Fisher information will approach 0 as $n \rightarrow \infty$. More flexible dependence structures for $\lambda$ and $\kappa$ could be accommodated a posteriori given different choices for $c(u_1, u_2; \boldsymbol{\upsilon})$. In those settings, eliciting dependence structures for $\lambda$ and $\kappa$ \color{black}{to support inference about posterior dependence} \color{black}{could} be worthwhile -- even for large samples since the dependence structure might be retained.

\subsection{Design Priors}

Not all prior distributions are elicited with the intent of defining a posterior. Prior distributions are regularly used for design purposes. For instance, priors might be used to summarize expert opinion to choose inputs for a decision model \citep{garthwaite2005statistical}. Design priors are also used in experimental settings to conduct sample size determination \citep{de2007using,berry2010bayesian,gubbiotti2011bayesian}. In these settings, design priors are often informative and concentrated on $\boldsymbol{\theta}$ values that are relevant to the objective of the study. Data generated according to the design prior are often combined with an uninformative analysis prior to assess whether a posterior criterion is satisfied. The prior $p_D(\boldsymbol{\theta})$ from (\ref{eq:prior_pred}) could be considered as a design prior in the context of this paper. 

Design priors are not directly combined with a likelihood function. It is therefore not an issue if the dependence structure in the \emph{design} prior satisfies the conditions for chronic rejection outlined in Corollary \ref{cor2}. Generally, we do not need to make separate considerations for small and large samples when using design priors. It is possible that the design prior might coincide with an analysis prior $p(\boldsymbol{\theta})$ that defines a posterior of $\boldsymbol{\theta}$. In that event, the design prior is subject to the previous recommendations in this section. The discussion on the coverage of credible sets in Section \ref{sec:design.cov} may be relevant if the design and analysis priors do not coincide.

\subsection{Posterior Concentration}

Recent work has suggested that the choice of prior dependence structure can expedite the convergence of the posterior around a fixed parameter value $\boldsymbol{\theta}_0$. \color{black}{This objective of increasing posterior concentration is related to the goal of improving model identifiability. The use of informative marginal priors to improve identifiability in complex settings has previously been discussed \citep{gelfand1999identifiability,eberly2000identifiability, gustafson2005model}.

In terms of the dependence structure,} \color{black}{Michimae} and Emura \citep{michimae2022bayesian} recently suggested that joint priors with vine structures based on Archimedean copulas lead to more accurate and concentrated posterior distributions in the context of Bayesian ridge regression. The accuracy and concentration of the posterior of  $\boldsymbol{\theta}$ were considered via the total mean absolute error between the posterior median and a fixed parameter value $\boldsymbol{\theta}_0$. Michimae and Emura's \citep{michimae2022bayesian} numerical studies showed that the posterior was more concentrated around $\boldsymbol{\theta}_0$ when the marginal priors for the regression coefficients were joined using (Archimedean) Clayton and Gumbel copulas \citep{nelsen2006introduction} instead of more standard Gaussian copulas.

\color{black}{Under the conditions for the Bernstein-von Mises theorem,} \color{black}{the} asymptotic theory from Section \ref{sec:posterior} suggests the choice of prior dependence structure cannot give rise to increased posterior concentration for large samples. For small samples, however, further investigation into how the prior dependence structure prompts increased posterior concentration is required. This investigation would disclose whether improving posterior concentration is a sensible objective for prior dependence specification. As such, we study the prior copula's impact on the posterior distribution in Section \ref{sec:analysis}. \color{black}{This} investigation considers the connection between posterior concentration and model identifiability and contextualizes our results within the broader discussion on the use of chronically rejected dependence structures.\color{black}




\section{Impact of the Prior Copula on the Posterior}\label{sec:analysis}

\subsection{Convergence of the Posterior Mode}

Since it may be challenging to correctly specify the dependence structure a priori, we consider how the choice of copula for the prior distribution impacts the posterior. \color{black}{We suppose that two potential priors for $\boldsymbol{\theta}$, denoted $p_1(\boldsymbol{\theta})$ and $p_2(\boldsymbol{\theta})$, are defined as in (\ref{eqn:prior}) using the \emph{same} marginal distributions $F_1, \dots, F_d$ but \emph{different} copula density functions $c_1(\boldsymbol{u})$ and $c_2(\boldsymbol{u})$.} \color{black}{We} require that both copula density functions are absolutely continuous and twice differentiable with respect to $\boldsymbol{u} = (u_1, \dots, u_d)$. To ensure the domain of the parameter space is not inadvertently restricted, the priors should be chosen such that $p_1(\boldsymbol{\theta}) > 0$ and $p_2(\boldsymbol{\theta}) > 0$ for all $\boldsymbol{\theta} \in \boldsymbol{\Theta}$. 

We define posteriors $p_1(\boldsymbol{\theta}\hspace{1pt}|\hspace{1pt} \boldsymbol{y}^{_{(n)}})$ and $p_2(\boldsymbol{\theta}\hspace{1pt}|\hspace{1pt} \boldsymbol{y}^{_{(n)}})$ by combining the likelihood $L(\boldsymbol{\theta};\hspace{1pt} \boldsymbol{y}^{_{(n)}})$ for the model \linebreak $m(\boldsymbol{y}^{_{(n)}} \hspace{1pt}| \hspace{1pt}\boldsymbol{\theta})$ with $p_1(\boldsymbol{\theta})$ and $p_2(\boldsymbol{\theta})$, respectively. We summarize each posterior via its posterior mode for $\boldsymbol{\theta}$, denoted by $\tilde{\boldsymbol{\theta}}^{_{(k)}} = \text{arg\,max}_{{\boldsymbol{\theta}}} ~p_k(\boldsymbol{\theta}\hspace{1pt}|\hspace{1pt} \boldsymbol{y}^{_{(n)}})$ for $k = 1,2$. In this section, we consider the convergence of the posterior mode to a fixed value $\boldsymbol{\theta}_0$ \color{black}{because the posterior generally concentrates around its mode as data are observed.} \color{black}{For} a given sample $\boldsymbol{y}^{_{(n)}}$, we compare the Euclidean distance between $\boldsymbol{\theta}_0$ and each posterior mode, denoted by $\mathcal{D}_k = \lVert \tilde{\boldsymbol{\theta}}^{_{(k)}} - \boldsymbol{\theta}_0 \rVert_2$ for $k = 1,2$. Theorem \ref{thm2} indicates that we generally do not expect the choice of prior copula to impact whether the posterior mode is closer to $\boldsymbol{\theta}_0$ for large sample sizes $n$.

 \begin{theorem}\label{thm2}
 Let $\boldsymbol{Y}^{_{(n)}}$ be generated independently from $m(\hspace{1pt} \cdot \hspace{1pt}| \hspace{1pt}\boldsymbol{\theta}_0)$ according to the Bernstein-von Mises theorem.  Let priors $p_1(\boldsymbol{\theta})$ and $p_2(\boldsymbol{\theta})$ be defined as in (\ref{eqn:prior}) with the same marginals $F_1, \dots, F_d$ but different copula density functions $c_1(\cdot)$ and $c_2(\cdot)$ that are absolutely continuous and twice differentiable. Suppose $p_1(\boldsymbol{\theta}) > 0$ and $p_2(\boldsymbol{\theta})> 0$ for all $\boldsymbol{\theta} \in \boldsymbol{\Theta}$. Given $\boldsymbol{y}^{_{(n)}}$, define posteriors $p_k(\boldsymbol{\theta}\hspace{1pt}|\hspace{1pt} \boldsymbol{y}^{_{(n)}}) \propto L(\boldsymbol{\theta};\hspace{1pt} \boldsymbol{y}^{_{(n)}}) \hspace{1pt} p_k(\boldsymbol{\theta})$ with posterior mode $\tilde{\boldsymbol{\theta}}^{_{(k)}}$ for $k = 1,2$. Let $\boldsymbol{u}_0 = (F_1(\theta_{1,0}), \dots , F_d(\theta_{d,0}))$ and $\mathcal{D}_k = \lVert \tilde{\boldsymbol{\theta}}^{_{(k)}} - \boldsymbol{\theta}_0 \rVert_2$ for $k = 1,2$.

 \begin{enumerate}
     \item[(a)] If $\nabla_{\boldsymbol{u}} \big[ \emph{log}(c_2(\boldsymbol{u})) - \emph{log}(c_1(\boldsymbol{u}))\big]_{\boldsymbol{u} = \boldsymbol{u}_0} \ne \boldsymbol{0}$, then $\lim_{n \rightarrow \infty} Pr(\mathcal{D}_2 \le \mathcal{D}_1) = 0.5$.
     \item[(b)] If $\boldsymbol{u}_0$ is a local maximum of $\emph{log}(c_2(\boldsymbol{u})) - \emph{log}(c_1(\boldsymbol{u}))$, then $\lim_{n \rightarrow \infty} Pr(\mathcal{D}_2 \le \mathcal{D}_1) = 1$.
 \end{enumerate}
	
\end{theorem}

Theorem \ref{thm2} explains that whether the choice of prior copula gives rise to faster convergence depends on the function $\text{log}(c_2(\boldsymbol{u})) - \text{log}(c_1(\boldsymbol{u}))$. \color{black}{The} posterior mode $\tilde{\boldsymbol{\theta}}^{_{(k)}}$ maximizes the logarithm of $p_k(\boldsymbol{\theta}\hspace{1pt}|\hspace{1pt} \boldsymbol{y}^{_{(n)}})$ for $k = 1, 2$. The log-posteriors $\text{log}\left[ p_1(\boldsymbol{\theta}\hspace{1pt}|\hspace{1pt} \boldsymbol{y}^{_{(n)}})\right]$ and $\text{log}\left[p_2(\boldsymbol{\theta}\hspace{1pt}|\hspace{1pt} \boldsymbol{y}^{_{(n)}})\right]$ differ \emph{only} by their prior copula log-density functions. Differences in $\tilde{\boldsymbol{\theta}}^{_{(1)}}$ and $\tilde{\boldsymbol{\theta}}^{_{(2)}}$ are therefore driven by differences in $\log(c_1(\boldsymbol{u}))$ and $\log(c_2(\boldsymbol{u}))$. Each copula log-density function prompts an additive contribution to the log-posterior. We suppose that $\boldsymbol{u}_0$ is not a stationary point of $\text{log}(c_2(\boldsymbol{u})) - \text{log}(c_1(\boldsymbol{u}))$ in part $(a)$. As $n$ increases, the contribution from $\log(c_2(\boldsymbol{u}))$ will not uniformly force its posterior mode closer to (or further from) $\boldsymbol{\theta}_0$ for all samples $\boldsymbol{y}^{_{(n)}}$ than that from $\log(c_1(\boldsymbol{u}))$. \color{black}{When} $\boldsymbol{u}_0$ is instead a local maximum of $\text{log}(c_2(\boldsymbol{u})) - \text{log}(c_1(\boldsymbol{u}))$, the contribution from $\log(c_2(\boldsymbol{u}))$ will force $\tilde{\boldsymbol{\theta}}^{_{(2)}}$ closer to $\boldsymbol{\theta}_0$ than $\tilde{\boldsymbol{\theta}}^{_{(1)}}$ for all samples $\boldsymbol{y}^{_{(n)}}$ as $n$ increases. This case is considered in part $(b)$. 

While \color{black}{$c_1(\cdot)$ and $c_2(\cdot)$ are functions of $\boldsymbol{\theta}$ because $\boldsymbol{u} = (F_1(\theta_1), \dots, F_d(\theta_d))$}\color{black}{,} we consider the partial derivatives of the copula log-density functions with respect to $\boldsymbol{u}$ instead of $\boldsymbol{\theta}$. For $j = 1, \dots, d$, it follows by the chain rule that

\vspace*{-10pt}
\begin{equation}\label{eq:deriv_cop}
\begin{split}
\frac{\partial}{\partial \theta_j} \bigg[ \text{log}(c_2(\boldsymbol{u})) - \text{log}(c_1(\boldsymbol{u}))\bigg] =  \\ f_j(\theta_j)\frac{\partial}{\partial u_j} \bigg[ \text{log}(c_2(\boldsymbol{u})) - \text{log}(c_1(\boldsymbol{u}))\bigg].
\end{split}
\end{equation}
We can factor out the $f_j(\theta_j)$ term because $p_1(\boldsymbol{\theta})$ and $p_2(\boldsymbol{\theta})$ are defined using the same marginals. The priors were also defined such that $p_1(\boldsymbol{\theta}) > 0$ and $p_2(\boldsymbol{\theta}) > 0$ for all $\boldsymbol{\theta} \in \boldsymbol{\Theta}$. Therefore, $f_j(\theta_j)$ must be positive, and the partial derivative in (\ref{eq:deriv_cop}) with respect to $\theta_j$ is 0 if and only if the partial derivative with respect to $u_j$ is 0. This correspondence ensures the results from Theorem \ref{thm2} generalize over different specifications for the marginal distributions \color{black}that are common to both priors. Counterparts to Theorem \ref{thm2} for settings where both the marginals and copulas are allowed to differ between $p_1(\boldsymbol{\theta})$ and $p_2(\boldsymbol{\theta})$ would be less broadly applicable and more conceptually complex.

\color{black}{We} prove Theorem \ref{thm2} in Appendix C of the online supplement. We note that if $\boldsymbol{u}_0$ is a local minimum of $\text{log}(c_2(\boldsymbol{u})) - \text{log}(c_1(\boldsymbol{u}))$, then $\lim_{n \rightarrow \infty} Pr(\mathcal{D}_2 \le \mathcal{D}_1) = 0$. This follows directly from part \emph{(b)} of Theorem \ref{thm2} by switching the labels on $c_1(\cdot)$ and $c_2(\cdot)$. The case where $\boldsymbol{u}_0$ is a saddle point of $\text{log}(c_2(\boldsymbol{u})) - \text{log}(c_1(\boldsymbol{u}))$ is excluded from both parts \emph{(a)} and \emph{(b)}. In that case, $Pr(\mathcal{D}_2 \le \mathcal{D}_1)$ may converge to a constant that is not 0.5 or 1. We explore the results from Theorem \ref{thm2} via simulation and explain their practical implications in Section \ref{sec:analysis.prac}.

\subsection{Practical Implications}\label{sec:analysis.prac}

Here, we conduct simulations to consider Theorem \ref{thm2} in practice. To do so, we consider an example adapted from Michimae and Emura \citep{michimae2022bayesian} since their recommendations are inconsistent with the results of Theorem \ref{thm2}. They considered a ridge regression model with three regression coefficients in the presence of multicollinearity, where the parameters of the relevant prior copulas were random variables specified using a hierarchical framework. The simplified example for our numerical study adapts aspects of their model for illustrative purposes. Note that we contrast our results with Michimae and Emura's \citep{michimae2022bayesian} findings in Section \ref{sec:analysis.rec}. 
 
Our simplified example considers the following linear regression model for the outcome $y_i$ and predictors $x_{i1}$ and $x_{i2}$:
\begin{equation*}
y_i = \beta_1x_{i1} + \beta_2x_{i2} + \varepsilon_i,
\end{equation*}
where $\varepsilon_i \sim \mathcal{N}(0, 5)$ independently for $i = 1, \dots, n$. The assumptions that the linear equation has an intercept of zero and the error terms have known variance reduce the dimensionality of the problem for illustration. That is, $\boldsymbol{\theta} = (\beta_1, \beta_2)$. We specify standard normal marginal priors for both $\beta_1$ and $\beta_2$. We join these marginal priors with two prior copulas in this numerical study: $c_1(\boldsymbol{u}) = 1$ for $\boldsymbol{u} \in [0,1]^2$ corresponds to the independence copula and $c_2(\boldsymbol{u})$ corresponds to a two-dimensional $t$-copula with $\nu = 4$ degrees of freedom and a diagonal correlation matrix $\boldsymbol{R}$ \color{black}(i.e, the $2\times2$ identity matrix). \color{black}Using the notation from Corollary \ref{cor2}, both copulas are such that $\tau^p_{\beta_1, \beta_2} = 0$. \color{black}

\begin{figure}[htb] \centering 
		\includegraphics[width = 0.475\textwidth]{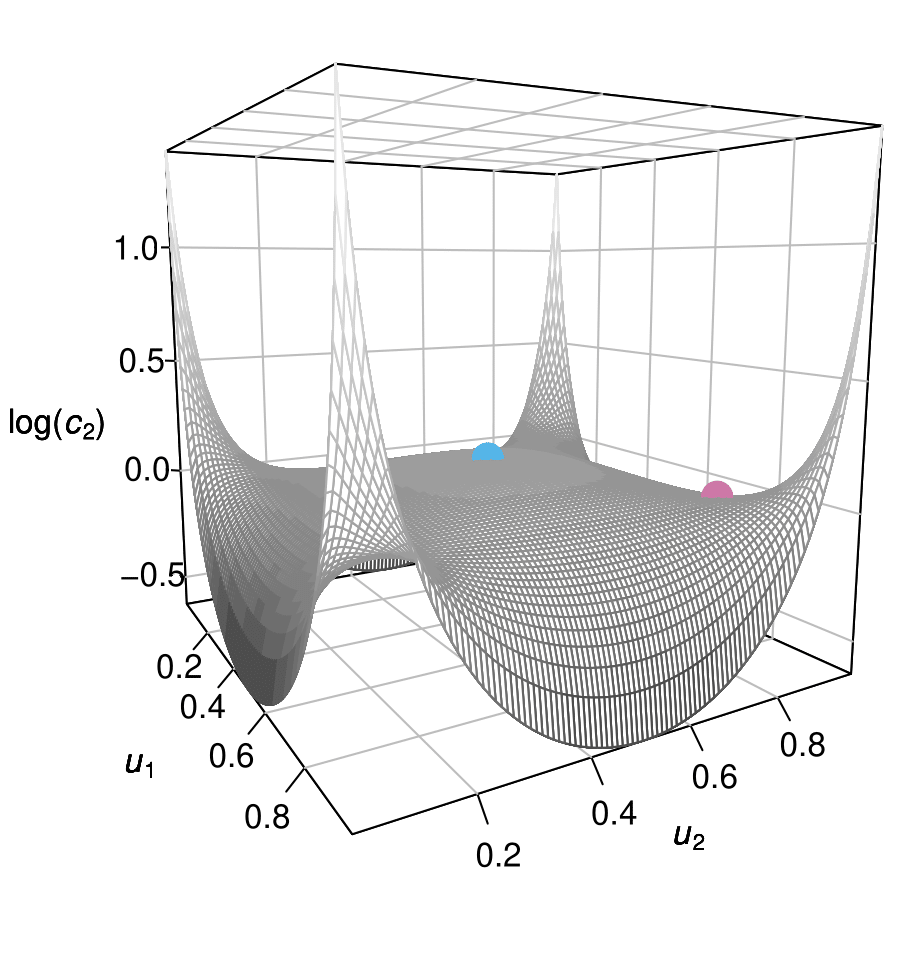} 
		\caption{\label{fig:t_copula} The logarithm of the $t$-copula density function with diagonal $\boldsymbol{R}$ and $\nu = 4$. The local maximum at $\boldsymbol{u} = (0.5, 0.5)$ and saddle point at $\boldsymbol{u} = (0.813, 0.813)$ are given by the blue and pink points, respectively.} 
	\end{figure}

We select the first copula because it is often assumed that the regression coefficients are independent a priori in Bayesian regression models. The corresponding joint prior for $\beta_1$ and $\beta_2$ is therefore a standard bivariate normal distribution with diagonal $\boldsymbol{R}$. The choice for the second copula is motivated by $c_2(\boldsymbol{u})$ having a local maximum and saddle points to illustrate the results from Theorem \ref{thm2}. Figure \ref{fig:t_copula} visualizes the logarithm of this copula density function. The selected $t$-copula does not accommodate strong negative or positive dependence between $\beta_1$ and $\beta_2$, but it reflects a greater likelihood of observing extreme values for both $\beta_1$ and $\beta_2$ relative to their marginal priors. For instance, this might occur if both marginal priors were misspecified.

  \begin{figure*}[!tb] \centering 
		\includegraphics[width = \textwidth]{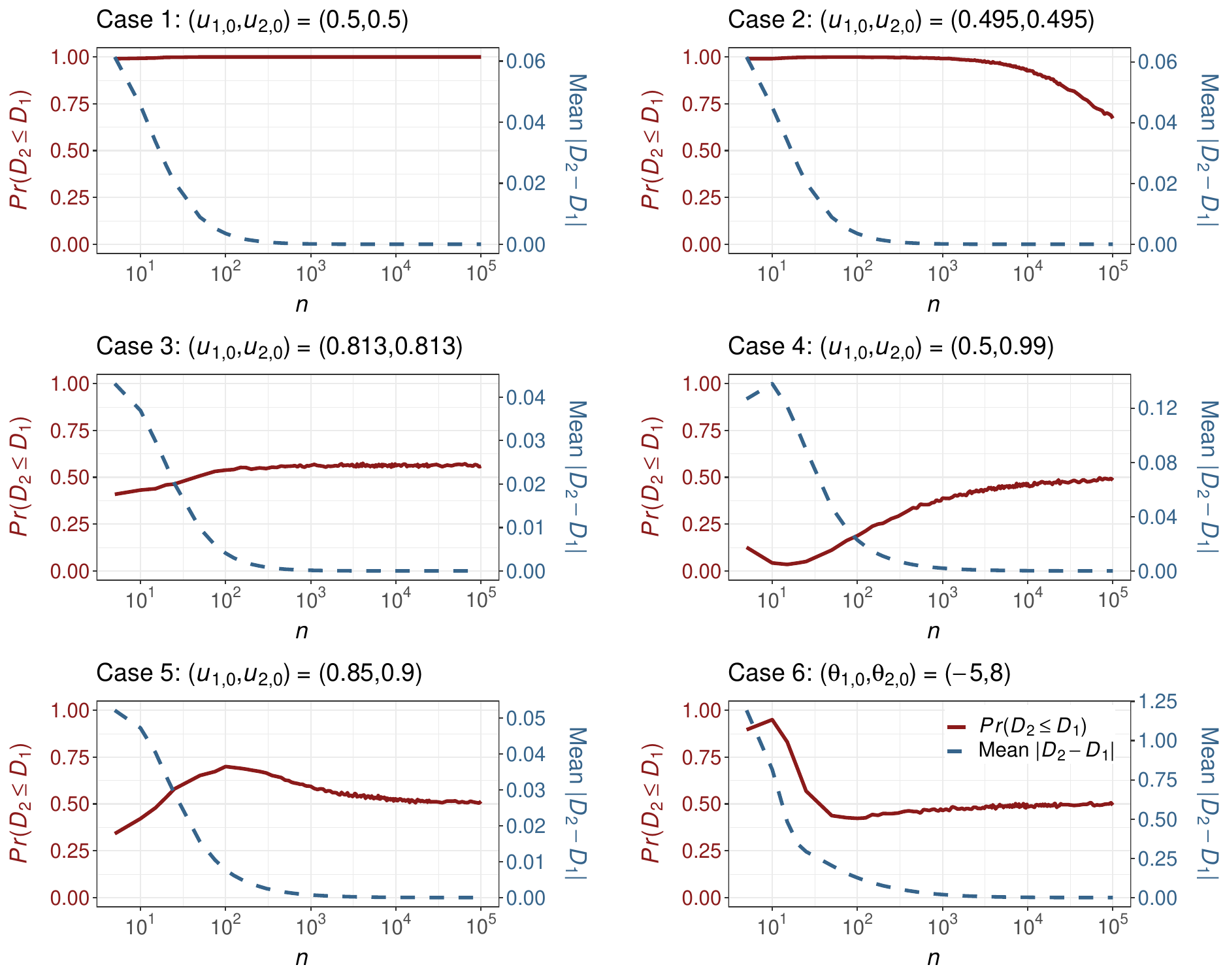} 
		\caption{\label{fig:thm2} Estimated probability that $\tilde{\boldsymbol{\theta}}^{_{(2)}}$ is closer to $\boldsymbol{\theta}_0$ than  $\tilde{\boldsymbol{\theta}}^{_{(1)}}$ (solid red) and mean absolute difference between $\mathcal{D}_2$ and $\mathcal{D}_1$ (dashed blue)  as a function of $n$ on the logarithmic scale (base 10) for six $\boldsymbol{\theta}_0$ values.} 
	\end{figure*} 

 We consider six values for $\boldsymbol{\theta}_0$ to define the data generation process for $\boldsymbol{Y}^{_{(n)}}$. These six values are meant to illustrate the convergence of the posterior mode in a variety of settings. Because $\boldsymbol{u}_0 = (F_1(\theta_{1,0}), F_2(\theta_{2,0}))$, we can readily convert between $\boldsymbol{\theta}_0$ and  $\boldsymbol{u}_0$ given the specified $\mathcal{N}(0,1)$ marginals. The marginal priors for $\beta_1$ and $\beta_2$ were chosen to be rather informative so that we observe a range of behaviour for the settings corresponding to part $(a)$ of Theorem \ref{thm2}.
 
 For each $\boldsymbol{\theta}_0$ value, we generated 10000 samples of size $n$ for various sample sizes between 5 and $10^5$. Each observation was simulated independently as $y_i = \beta_{1,0}x_{i1} + \beta_{2,0}x_{i2} + \varepsilon_i$, where $(x_{i1}, x_{i2}) \sim \mathcal{N}(\boldsymbol{0}, I_2)$ and $\varepsilon_i \sim \mathcal{N}(0,5)$ for $i = 1,\dots,n$. \color{black}This data generation process implies $\tau(\boldsymbol{\theta}_0)_{\beta_1, \beta_2} = 0$. \color{black}For each of these 10000 samples, we found both posterior modes $\tilde{\boldsymbol{\theta}}^{_{(1)}}$ and $\tilde{\boldsymbol{\theta}}^{_{(2)}}$ \color{black}{by numerically finding the zeros of the partial derivatives.} \color{black}{We} then estimated $Pr(\mathcal{D}_2 \le \mathcal{D}_1)$ as the proportion of samples for which $\tilde{\boldsymbol{\theta}}^{_{(2)}}$, corresponding to the $t$-copula, was closer to $\boldsymbol{\theta}_0$ than  $\tilde{\boldsymbol{\theta}}^{_{(1)}}$. To consider the practical impact of Theorem \ref{thm2}, we also computed the mean absolute difference between $\mathcal{D}_2$ and $\mathcal{D}_1$ for each sample size $n$. Figure \ref{fig:thm2} visualizes these results for each of the six scenarios, which we now describe. 

We first consider case 1, where $\boldsymbol{u}_0$ = (0.5, 0.5). This point is a local maximum of $c_2(\boldsymbol{u})$ and therefore also a local maximum of $\text{log}(c_2(\boldsymbol{u})) - \text{log}(c_1(\boldsymbol{u})) = \text{log}(c_2(\boldsymbol{u}))$. As indicated by Theorem \ref{thm2}, the probability that $\tilde{\boldsymbol{\theta}}^{_{(2)}}$ is closer to $\boldsymbol{\theta}_0$ than  $\tilde{\boldsymbol{\theta}}^{_{(1)}}$ approaches 1 as $n \rightarrow \infty$. Even though $\mathcal{D}_2$ is less than  $\mathcal{D}_1$, their absolute difference is practically negligible for large sample sizes $n$. Case 2 considers a point $\boldsymbol{u}_0$ = (0.495, 0.495) that is extremely close to the local maximum. The plot for case 2 is similar to the previous one for small $n$, but the estimate for $Pr(\mathcal{D}_2 \le \mathcal{D}_1)$ slowly decreases as $n \rightarrow \infty$. The estimated probability is still 0.673 for $n = 10^5$. For $n = 5\times10^6$ (not pictured), we estimated $Pr(\mathcal{D}_2 \le \mathcal{D}_1)$ to be 0.521. While $Pr(\mathcal{D}_2 \le \mathcal{D}_1)$ approaches 0.5 in theory, it may not be 0.5 in practice if the sample size must be prohibitively large for the asymptotic result to hold. Case 3 examines a saddle point at $\boldsymbol{u}_0$ = ($\tilde{F}(1; 4)$, $\tilde{F}(1; 4)$), where $\tilde{F}(\cdot; 4)$ is the CDF of the Student's $t$-distribution with $\nu = 4$ and $\tilde{F}(1; 4) \approx 0.813$. This setting is noteworthy because $Pr(\mathcal{D}_2 \le \mathcal{D}_1)$ does not approach 0.5 or 1. The probability for this scenario instead approaches 0.563, and we confirmed this limiting probability using samples of size $n = 5\times10^6$. 
 

\begin{table*}[tb]
\caption{Empirical coverage of 95\% HPD sets for $\boldsymbol{\theta} = (\beta_{1}, \beta_{2})$ across 10000 posteriors defined using both prior copulas. \color{black}{The median area of the HPD set across the 10000 posteriors for each case and prior combination is given in parentheses.}}
\label{tab:cov}
\begin{tabular}{@{}lcccccc@{}}
\hline
& \multicolumn{6}{c}{Case for Independence Copula} \\[1pt]
\cline{2-7} \\[-6.1pt]
$n$ & \multicolumn{1}{c}{1} & \multicolumn{1}{c}{2}
& \multicolumn{1}{c}{3} & \multicolumn{1}{c}{4} & \multicolumn{1}{c}{5} & \multicolumn{1}{c}{6} \\
\hline
$10^1$ & 0.9928 (6.8480) & 0.9916 (6.8347) & 0.9685 (6.8451) & 0.8723 (6.8440) & 0.9465 (6.8627) & 0.0023 (6.8552)

 \\
        $10^2$ & 0.9574 (0.9260) & 0.9575 (0.9261) & 0.9558 (0.9281) & 0.9387 (0.9264) & 0.9479 (0.9258) & 0.5698 (0.9285)

 \\
        $10^3$ & 0.9510 (0.0957) & 0.9550 (0.0957) & 0.9515 (0.0957) & 0.9509 (0.0957) & 0.9504 (0.0957) & 0.9181 (0.0958)

 \\
        $10^4$ & 0.9519 (0.0096) & 0.9525 (0.0096) & 0.9517 (0.0096) & 0.9546 (0.0096) & 0.9542 (0.0096) & 0.9452 (0.0096)

 \\
        $10^5$ & 0.9546 (0.0010) & 0.9501 (0.0010) & 0.9507 (0.0010) & 0.9523 (0.0010) & 0.9473 (0.0010) & 0.9508 (0.0010)

 \\ \hline
 & \multicolumn{6}{c}{Case for $t$-Copula} \\[1pt]
\cline{2-7} \\[-6.1pt]
$n$ & \multicolumn{1}{c}{1} & \multicolumn{1}{c}{2}
& \multicolumn{1}{c}{3} & \multicolumn{1}{c}{4} & \multicolumn{1}{c}{5} & \multicolumn{1}{c}{6} \\
\hline
$10^1$ & 0.9939 (6.6420) & 0.9938 (6.6201) & 0.9683 (6.7271) & 0.7769 (6.9754) & 0.9465 (6.7904) & 0.0002 (4.7521)

 \\
        $10^2$ & 0.9583 (0.9157) & 0.9603 (0.9159) & 0.9570 (0.9238) & 0.9234 (0.9413) & 0.9503 (0.9220) & 0.5148 (0.9482)
 \\
        $10^3$ & 0.9509 (0.0956) & 0.9558 (0.0956) & 0.9532 (0.0957) & 0.9481 (0.0959) & 0.9507 (0.0957) & 0.9162 (0.0960)

 \\
        $10^4$ & 0.9515 (0.0096) & 0.9532 (0.0096) & 0.9524 (0.0096) & 0.9512 (0.0096) & 0.9553 (0.0096) & 0.9451 (0.0096)

 \\
        $10^5$ & 0.9539 (0.0010) & 0.9503 (0.0010) & 0.9518 (0.0010) & 0.9516 (0.0010) & 0.9479 (0.0010) & 0.9524 (0.0010)

 \\ \hline
\end{tabular}
\end{table*}

 Case 4 considers a point $\boldsymbol{u}_0$ = (0.5, 0.99) where the independence copula performs better for smaller sample sizes in that the estimated probability approaches 0.5 from below. We note that $c_2(\boldsymbol{u}_0) = 0.533 < 1 = c_1(\boldsymbol{u}_0)$, which occurs because this bivariate $t$-copula deems scenarios where only one of $\beta_1$ or $\beta_2$ is extreme relative to their marginal priors as more rare than the independence copula. If $c_2(\boldsymbol{u}_0)$ is less (greater) than $c_1(\boldsymbol{u}_0)$, $Pr(\mathcal{D}_2 \le \mathcal{D}_1)$ often approaches 0.5 from below (above). However, this behaviour is not guaranteed. Case 5 examines a point $\boldsymbol{u}_0$ = (0.85, 0.9) at which the estimated probability approaches 0.5 from above. Here, $c_2(\boldsymbol{u}_0) = 1.047 > 1$, but the estimate for $Pr(\mathcal{D}_2 \le \mathcal{D}_1)$ is less than 0.5 for sample sizes less than $n = 20$. Lastly, case 6 is defined in terms of its value for $\boldsymbol{\theta}_0 = (\beta_{1,0}, \beta_{2,0}) = (-5, 8)$ because the corresponding $\boldsymbol{u}_0$ value of ($2.87 \times 10^{-7}$, $1 - 6.22\times10^{-16}$) is quite extreme. For this setting, $c_2(\boldsymbol{u}_0) = 5054.68 >> 1$, and $\mathcal{D}_2$ is roughly 0.8 smaller than $\mathcal{D}_1$ on average for $n = 10$. Yet, the estimated probability approaches 0.5 from below for large sample sizes. This likely occurs because $c_2(\boldsymbol{u})$ is volatile near $\boldsymbol{u}_0$.  

 For each case and prior combination, we also estimated empirical coverage as the proportion of the 10000 posteriors for which the 95\% HPD set included the parameter value $\boldsymbol{\theta}_0$ when $n = \{10^1, 10^2, 10^3, 10^4, 10^5 \}$. The HPD sets were again estimated using two-dimensional kernel density estimation on posterior draws obtained via sampling-resampling methods. \color{black}We estimated the median area of the 10000 HPD sets for each case and prior combination using quasi-Monte Carlo methods \citep{lemieux2009using}. \color{black}{The} empirical coverage \color{black} and median area \color{black}{results} are summarized for both prior copulas in Table \ref{tab:cov}. 

Per Table \ref{tab:cov}, the empirical coverage is similar for both prior copulas at all samples sizes considered in cases 1, 2, 3, and 5. For small sample sizes, the empirical coverage exceeds the nominal value of 95\% in cases 1, 2, and 3. Since this trend is observed for both prior copulas, it is caused by the informative nature of the relatively well-specified marginal priors for $\beta_1$ and $\beta_2$. These marginal priors are misspecified in cases 4 and 6, so the empirical coverage is less than 95\% for both prior copulas when the sample size is small. The empirical coverage is better when using the independence copula for small $n$ in both cases -- even though $\tilde{\boldsymbol{\theta}}^{_{(2)}}$ was generally closer to $\boldsymbol{\theta}_0$ than $\tilde{\boldsymbol{\theta}}^{_{(1)}}$ for case 6 when $n$ was small in Figure \ref{fig:thm2}. 

\color{black}
Neither prior copula reliably prompts greater posterior concentration \emph{and} coverage. The $t$-copula gives rise to slightly smaller median areas of the 95\% HPD sets for cases 1, 2, 3, and 5 when $n = 10^1$ or $10^2$. However, the independence copula yields smaller median areas at these sample sizes for case 4. In case 6, the median area is much smaller for the $t$-copula when $n = 10^1$, but this increase in posterior concentration is accompanied by a decrease in coverage. The independence copula prompts both greater coverage and posterior concentration for case 6 when $n = 10^2$. 
\color{black}

\subsection{Connections to Other Work}\label{sec:analysis.rec}

 Our simulations demonstrate the value in considering the copula density functions $c_1(\boldsymbol{u})$ and $c_2(\boldsymbol{u})$ when choosing between two prior dependence structures. In particular, we may want to consider the local optima for the copula densities. This numerical study also suggests that copulas have limited ability to reliably prompt more accurate and concentrated posterior distributions around a fixed parameter value $\boldsymbol{\theta}_0$. \color{black}The $t$-copula only reliably prompted a more accurate and concentrated posterior when $\boldsymbol{u}_0$ was in a neighbourhood of (0.5, 0.5). To reap these benefits, we would need to know that $\boldsymbol{u}_0$ is in a particular region of $[0,1]^d$ a priori. In that event, we should likely make our marginal priors more informative to reflect this knowledge. Such strategies would be more related to past work on model identifiability \citep{gelfand1999identifiability,eberly2000identifiability, gustafson2005model}.

 \color{black}
 In practice, the true value of $\boldsymbol{u}_0$ is instead unknown. And even \emph{if} $\boldsymbol{u}_0$ is such that $c_2(\boldsymbol{u}_0) > c_1(\boldsymbol{u}_0)$, our numerical studies demonstrate that $\tilde{\boldsymbol{\theta}}^{_{(2)}}$ is not guaranteed to be closer to $\boldsymbol{\theta}_0$ than $\tilde{\boldsymbol{\theta}}^{_{(1)}}$ for small or large sample sizes. \color{black}Both copula priors in our numerical study were such that $\tau^p_{\beta_1, \beta_2} = \tau(\boldsymbol{\theta}_0) = 0$; neither prior dependence structure satisfied the conditions for Corollary \ref{cor2}. Regardless of whether the prior dependence structure is a chronically rejected one, choosing a prior copula to improve posterior concentration is likely not a sensible objective.

 \color{black}
 We now contrast our conclusions with Michimae and Emura's \citep{michimae2022bayesian} recommendations. Their numerical studies considered several levels of multicollinearity between the regression covariates for a \emph{single} $\boldsymbol{\theta}_0$ value.  The joint prior for their three regression coefficients leveraged a vine structure with three bivariate copulas. For each bivariate copula, the corresponding $\boldsymbol{u}_0$ value lies directly on the upper Fréchet-Hoeffding bound. Each $\boldsymbol{u}_0$ value does not necessarily correspond to a local maximum of the Clayton or Gumbel copula density functions they considered. However, the Clayton and Gumbel families of copulas accommodate positive prior dependence and their density functions are generally largest near the upper Fréchet-Hoeffding bound. Their single $\boldsymbol{\theta}_0$ value is therefore similar to that used in case 2 of our study. They also considered relatively small sample sizes ranging from $n = 20$ to 200. Given the results for case 2 in Figure \ref{fig:thm2}, it is not surprising that their normal posterior medians for the regression coefficients were generally closer to their $\boldsymbol{\theta}_0$ values when using Clayton and Gumbel copulas, \color{black}despite this result not holding more generally. \color{black}Our numerical results suggest that it is pertinent to consider a variety of $\boldsymbol{\theta}_0$ values and sample sizes $n$ before making general statements about the impact of the prior copula on the posterior distribution. 
 
 We acknowledge that Michimae and Emura \citep{michimae2022bayesian} considered a more complicated model, which leveraged a hierarchical framework to specify the prior copula. Although not the focus of this article, Monte Carlo simulation could likely be used to explore the local optima of such prior copula density functions and extend the results from Theorem \ref{thm2} to a hierarchical framework. This extension is one of several possible generalizations that could be made to the definition of the analysis prior in (\ref{eqn:prior}).   

\section{Discussion}\label{sec:disc}

    In this paper, we proposed a framework to consider whether prior dependence structures can be retained a posteriori. This framework \color{black} informs conversations between statisticians and practitioners to promote alignment between the flexibility in the prior dependence structure and the objectives for posterior analysis. Discarding chronically rejected dependence structures that cannot be retained as data are observed simplifies the prior specification process when practitioners aim to collect large samples or prioritize inference on the dependence structure. We also discussed small-sample objectives for prior specification to clarify that the inability to retain the prior dependence structure may not present practical issues for various posterior analyses. \color{black}This paper emphasized copula-based priors, but the notion of chronically rejected dependence structures is still applicable to multivariate prior distributions that are not explicitly defined using copulas, such as multivariate normal priors. 

    
    Since correctly specifying the dependence structure a priori for an arbitrary model may be difficult, we examined how the choice of prior copula impacts the posterior distribution. We proved asymptotic results regarding how this choice of copula impacts the convergence of the posterior mode to a fixed parameter value $\boldsymbol{\theta}_0$. While examining the local optima of candidate copula density functions is valuable, our numerical studies showed that prior copulas should not be selected to improve the posterior's ability to recover $\boldsymbol{\theta}_0$. These results contradicted past recommendations that suggested the choice of prior copula could reliably improve posterior concentration.

    Future work could extend the results from this paper to hierarchical settings, in which the hyperparameters of the prior copula are themselves random variables. Moreover, the theoretical results in this paper are based on the standard Bernstein-von Mises theorem. To broaden the applicability of this framework, we could consider nonparametric methods for specifying prior dependence. We could also consider chronically rejected prior dependence structures in the presence of model misspecification (i.e., when $L(\boldsymbol{\theta};\hspace{1pt} \boldsymbol{y})$ that defines the posterior does not coincide with $m(\hspace{1pt} \cdot \hspace{1pt}| \hspace{1pt}\boldsymbol{\theta})$ used to generate the data). In that case, we require more complex characterizations of prior dependence than $\mathcal{I}(\boldsymbol{\theta}_0)^{-1}$ for large samples.

\section*{Supplementary Material}
Appendix A proves the Fisher information for the conditional parameterization of multinomial model is diagonal. Appendix B contains additional simulations for the calibration of credible sets. Appendix C proves Theorem \ref{thm2}. The code for the numerical studies is provided on Github: \url{https://github.com/lmhagar/PosteriorRamifications}

	\section*{Funding Acknowledgement}
	This work was supported by the Natural Sciences and Engineering Research Council of Canada (NSERC) by way of a PGS D scholarship as well as Grant RGPIN-2019-04212.
	
	
\bibliographystyle{chicago}

\end{document}


\newcommand{\bb}{\boldsymbol{\beta}}

	\title{Posterior Ramifications of Prior Dependence Structures \medskip\\
 \Large{Supplementary Material}}


	\author{Luke Hagar \hspace{35pt} Nathaniel T. Stevens}

	\date{}

	\maketitle

	\maketitle

	\baselineskip=19.5pt


\appendix
\numberwithin{equation}{section}
\renewcommand{\theequation}{\thesection.\arabic{equation}}

\numberwithin{figure}{section}
\renewcommand{\thefigure}{\thesection.\arabic{figure}}

\numberwithin{table}{section}
\renewcommand{\thetable}{\thesection.\arabic{table}}

 \section{Fisher Information for the Conditional Multinomial Model}\label{sec:empir}

\begin{lemma}\label{lem1}
Let $\boldsymbol{\theta} = (Z_1, Z_2, ..., Z_{w-1})$ be the conditional probabilities defined in (2) of the main text for the standard multinomial model with $w$ categories. Then, the inverse Fisher information matrix $\mathcal{I}(\boldsymbol{\theta})^{-1}$ is diagonal for all possible $(Z_1, Z_2, ..., Z_{w-1}) \in (0,1)^{w-1}$.
\end{lemma}

To prove Lemma \ref{lem1}, we consider fixed parameters $\{\boldsymbol{p}_{w-1,0} = (p_{1,0}, ..., p_{w,0}) : 0 < p_{1,0}, ..., p_{w,0} < 1, \linebreak \sum_{v=1}^w p_{v,0} = 1\}$. The inverse of the Fisher information $\mathcal{I}(\boldsymbol{p}_{w-1,0})$ is given by $\mathcal{I}(\boldsymbol{p}_{w-1,0})^{-1} = n^{-1}\boldsymbol{\Sigma}$, where $\boldsymbol{\Sigma}_{s,t} = p_{s,0}(1-p_{s,0})$ if $1 \le s = t \le w-1$ and $-p_{s,0}p_{t,0}$ otherwise. We let $\hat{p}_v$ and $\hat{p}_{v^{_{\prime}}}$ for $1 \le v < v^{_{\prime}} \le w-1$ be MLEs for the multinomial parameters. We denote the vector of all such MLEs as $\hat{\boldsymbol{p}}_{w-1} = (\hat{p}_1, ..., \hat{p}_{w-1})$. We define analogues to the transformations in (2) of the main text for the fixed parameter values as $(Z_{1,0}, Z_{2,0}, ..., Z_{w-1,0})$.

 We let $Z_{v,0} = g_v(\boldsymbol{p}_{w-1,0})$ for $v = 1, ..., w-1$, and define $\hat{Z}_{v} = g_v(\hat{\boldsymbol{p}}_{w-1})$. By the multivariate delta method, the asymptotic covariance $ACov(\cdot, \cdot)$ between $g_v(\hat{\boldsymbol{p}}_{w-1})$ and $g_{v^{_{\prime}}}(\hat{\boldsymbol{p}}_{w-1})$ for $1 \le v < v^{_{\prime}} \le w-1$ is
 \begin{equation}\label{eqn:approx_cov}
		ACov(g_v(\hat{\boldsymbol{p}}_{w-1}), g_{v^{_{\prime}}}(\hat{\boldsymbol{p}}_{w-1}))  = \dfrac{1}{n} \sum_{s=1}^{w-1} \sum_{t=1}^{w-1} \dfrac{\partial g_v}{\partial p_{s,0}} \dfrac{\partial g_{v^{_{\prime}}}}{\partial p_{t,0}} \boldsymbol{\Sigma}_{s,t}.
	\end{equation} 
We use induction to show that $ACov(g_v(\hat{\boldsymbol{p}}_{w-1}), g_{v^{_{\prime}}}(\hat{\boldsymbol{p}}_{w-1})) = 0$ for all $1 \le v < v^{_{\prime}} \le w-1$. We first show that  $ACov(g_1(\hat{\boldsymbol{p}}_{2}), g_{2}(\hat{\boldsymbol{p}}_{2})) = 0$. This base case corresponds to the model with $w = 3$. We have that 
 \begin{equation*}\label{eqn:ord_deriv2}
		  \begin{aligned}
    \dfrac{\partial g_1}{\partial \boldsymbol{p}_{2,0}} &= \begin{bmatrix}
           1 \vspace*{2pt} \\
           0 
         \end{bmatrix}
  \end{aligned} ~~~~\text{and}~~~~
   \begin{aligned}
    \dfrac{\partial g_2}{\partial \boldsymbol{p}_{2,0}} &= \dfrac{p_{2,0}}{(1 - p_{1,0})^2}\begin{bmatrix}
           1 \vspace*{2pt} \\
           (1 - p_{1,0})/p_{2,0} 
         \end{bmatrix}.
  \end{aligned} 
	\end{equation*} 
Because $\partial g_1/ \partial p_{2,0} = 0$, it follows that $ACov(\hat{Z}_{1}, \hat{Z}_{2}) = 0$ for the base case:
  \begin{equation}\label{eqn:ord_deriv2p2}
 \begin{aligned}
n \times ACov(g_1(\hat{\boldsymbol{p}}_{2}), g_{2}(\hat{\boldsymbol{p}}_{2})) &= \dfrac{\partial g_1}{\partial p_{1,0}} \dfrac{\partial g_{2}}{\partial p_{1,0}} \boldsymbol{\Sigma}_{1,1} + \dfrac{\partial g_1}{\partial p_{1,0}} \dfrac{\partial g_{2}}{\partial p_{2,0}} \boldsymbol{\Sigma}_{1,2}  \\[2pt] 
       &=\dfrac{p_{1,0}p_{2,0}}{(1 - p_{1,0})^2}\left[1 - p_{1,0} - p_{2,0}(1 - p_{1,0})/p_{2,0}\right]\\
       &=0.
\end{aligned}
\end{equation} 

For the inductive hypothesis, we assume that $ACov(g_v(\hat{\boldsymbol{p}}_{k-1}), g_{v^{_{\prime}}}(\hat{\boldsymbol{p}}_{k-1})) = 0$ for all $1 \le v < v^{_{\prime}} \le k-1$. This implies that the result for the base case holds for an arbitrary multinomial model with $w = k$ categories. For the inductive conclusion, we show that this result also holds for an arbitrary model with $w = k + 1$ categories. With $k + 1$ categories, we have that 
\begin{equation*}\label{eqn:ord_derivk}
		  \begin{aligned}
    \dfrac{\partial g_1}{\partial \boldsymbol{p}_{k,0}} &= \begin{bmatrix}
           1 \vspace*{2pt} \\
           \boldsymbol{0}_{k-1} 
         \end{bmatrix},
  \end{aligned} 
   \begin{aligned}
    \dfrac{\partial g_v}{\partial \boldsymbol{p}_{k,0}} &= \dfrac{p_{v,0}}{\left(1 - \sum_{t = 1}^{v-1} p_{t,0}\right)^2}\begin{bmatrix}
           \boldsymbol{1}_{v-1} \vspace*{2pt} \\
           \left(1 - \sum_{t = 1}^{v-1} p_{t,0}\right)/p_{v,0} \vspace*{2pt} \\
           \boldsymbol{0}_{k-v}
         \end{bmatrix}, ~\text{for}~ v = 2, ..., k-1,
  \end{aligned} 
	\end{equation*} 
 
 \begin{equation}\label{eqn:ord_derivk2}
  \begin{aligned}
   \text{and}~~~ \dfrac{\partial g_k}{\partial \boldsymbol{p}_{k,0}} &= \dfrac{p_{k,0}}{\left(1 - \sum_{t = 1}^{k-1} p_{t,0}\right)^2}\begin{bmatrix}
           \boldsymbol{1}_{k-1} \vspace*{2pt} \\
           \left(1 - \sum_{t = 1}^{k-1} p_{t,0}\right)/p_{k,0}
         \end{bmatrix}.
  \end{aligned} 
	\end{equation} 
 Because of the upper triangular structure of the partial derivatives in (\ref{eqn:ord_derivk2}) and the inductive hypothesis, $ACov(g_v(\hat{\boldsymbol{p}}_{k}), g_{v^{_{\prime}}}(\hat{\boldsymbol{p}}_{k})) = 0$ for all $1 \le v < v^{_{\prime}} \le k-1$. We therefore just need to consider $ACov(g_v(\hat{\boldsymbol{p}}_{k}), g_k(\hat{\boldsymbol{p}}_{k}))$ for all $1 \le v \le k-1$. We now derive general expressions for these asymptotic covariances.

 We first derive the expression for $ACov(g_1(\hat{\boldsymbol{p}}_{k}), g_k(\hat{\boldsymbol{p}}_{k}))$. Similar to (\ref{eqn:ord_deriv2p2}), we find
  \begin{equation*}\label{eqn:ord_derivkp2}
 \begin{aligned}
n \times ACov(g_1(\hat{\boldsymbol{p}}_{k}), g_{k}(\hat{\boldsymbol{p}}_{k})) &= \sum_{s = 1}^{k} \dfrac{\partial g_1}{\partial p_{1,0}} \dfrac{\partial g_{k}}{\partial p_{s,0}} \boldsymbol{\Sigma}_{1,s}  \\[2pt] 
       &=\dfrac{p_{1,0}p_{k,0}}{\left(1 - \sum_{t = 1}^{k-1} p_{t,0}\right)^2}\left[\left(1 - \sum_{t = 1}^{k-1} p_{t,0}\right) - p_{k,0}\left(1 - \sum_{t = 1}^{k-1} p_{t,0}\right)/p_{k,0}\right]\\
       &=0
\end{aligned}
\end{equation*}
We now find an expression for $ACov(g_v(\hat{\boldsymbol{p}}_{k}), g_k(\hat{\boldsymbol{p}}_{k}))$, $v = 2, ..., k - 1$. By matrix multiplication, (\ref{eqn:ord_derivk2}), and the expression for $\boldsymbol{\Sigma}$, we obtain 
  \begin{equation*}\label{eqn:ord_derivkp3}
 \begin{aligned}
n \times ACov(g_v(\hat{\boldsymbol{p}}_{k}), g_{k}(\hat{\boldsymbol{p}}_{k})) &= \sum_{s = 1}^{v} \sum_{t = 1}^{k} \dfrac{\partial g_v}{\partial p_{s,0}} \dfrac{\partial g_{k}}{\partial p_{t,0}} \boldsymbol{\Sigma}_{s,t}  \\[2pt] 
       &=\dfrac{p_{v,0}p_{k,0}}{\left(1 - \sum_{t = 1}^{v-1} p_{t,0}\right)^2\left(1 - \sum_{t = 1}^{k-1} p_{t,0}\right)^2} \\
       & ~~~\times  \Bigg\{ \sum_{s=1}^{v-1}p_{s,0}\left[\left(1 - \sum_{t = 1}^{k-1} p_{t,0}\right) - p_{k,0}\left(1 - \sum_{t = 1}^{k-1} p_{t,0}\right)/p_{k,0}\right] +  \\[2pt] 
       & ~~~~~~~  \left(1 - \sum_{s=1}^{v-1}p_{s,0}\right)\left[\left(1 - \sum_{t = 1}^{k-1} p_{t,0}\right) - p_{k,0}\left(1 - \sum_{t = 1}^{k-1} p_{t,0}\right)/p_{k,0}\right] \Bigg\}
      \\
       &=0
\end{aligned}
\end{equation*}
Therefore, $ACov(g_v(\hat{\boldsymbol{p}}_{k}), g_{v^{_{\prime}}}(\hat{\boldsymbol{p}}_{k})) = 0$ for all $1 \le v < v^{_{\prime}} \le k$. This result holds for arbitrary $\boldsymbol{p}_{w-1,0}$. By mathematical induction, $ACov(\hat{Z}_v, \hat{Z}_{v^{_{\prime}}}) = 0$ for all $1 \le v < v^{_{\prime}} \le w-1$ for an arbitrary multinomial model with $\{w \in \mathbb{N}: w \ge 3\}$ categories. Therefore, $\mathcal{I}(\boldsymbol{\theta})^{-1}$ is diagonal for all possible $(Z_1, Z_2, ..., Z_{w-1}) \in (0,1)^{w-1}$. 

\section{Additional Simulations for the Calibration of Credible Sets}\label{sec:appendix.cov} 

Here, we investigate how the prior dependence structure impacts the calibration of Bayesian credible sets for an example where the prior dependence structure is not a chronically rejected one. We consider the standard gamma model parameterized by $\boldsymbol{\theta} = (\alpha, \beta)$, where $\alpha$ and $\beta$ are respectively the shape and rate parameters. The correlation between $\alpha$ and $\beta$ dictated by the inverse Fisher information matrix is $1/\sqrt{\alpha \psi_1(\alpha)}$, where $\psi_1(\cdot)$  is the trigamma function. The correlation $1/\sqrt{\alpha \psi_1(\alpha)}$ is a positive and increasing function for all $\alpha > 0$. When the conditions for Theorem 1 of the main text hold, the joint posterior of $\boldsymbol{\theta}$ will be unable to retain negative dependence structures between $\alpha$ and $\beta$. However, positive dependence structures between $\alpha$ and $\beta$ do not satisfy the conditions for chronic rejection outlined in Corollary 1 -- even if the magnitude of the prior dependence is not retained a posteriori.

\begin{figure}[!b] \centering 
		\includegraphics[width = 0.55\textwidth]{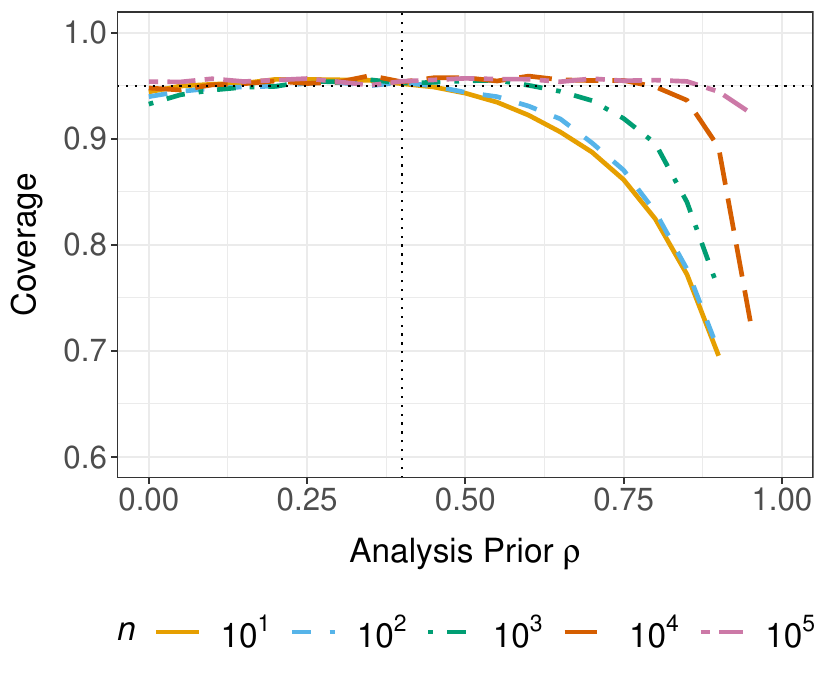} 
		\caption{\label{fig:cov} Empirical coverage of 95\% HPD for the gamma parameter $\boldsymbol{\theta} = (\alpha, \beta)$ across 10000 posteriors. The horizontal dotted line denotes the nominal coverage, and the vertical one denotes ``nature's'' prior.} 
	\end{figure}

We define the prior predictive distribution of $\boldsymbol{Y}^{_{(n)}}$ for these simulations by joining a $\text{GAMMA}(1000,5000)$ prior for $\alpha$ and a $\text{GAMMA}(1000,800)$ prior for $\beta$ with a Gaussian copula parameterized with Pearson's $\rho = 0.4$. For each of 10000 simulation repetitions, we approximated the posterior of $\boldsymbol{\theta} \hspace{1pt} | \hspace{1pt} \boldsymbol{y}^{_{(n)}}$ using sampling-resampling methods \citep{rubin1988using} with ``nature's'' analysis prior. The proposal distribution was the posterior of $\boldsymbol{\theta} \hspace{1pt} | \hspace{1pt} \boldsymbol{y}^{_{(n)}}$ obtained by independently joining the marginal priors for $\alpha$ and $\beta$, and we sampled from the proposal distribution using Markov chain Monte Carlo methods. For each posterior, we approximated its 95\% highest posterior density (HPD) set for $\alpha$ and $\beta$ using two-dimensional kernel density estimation \citep{ripley2002modern}. Empirical coverage was estimated as the proportion of simulation repetitions for which the parameter value $\boldsymbol{\theta}_0 = (\alpha_{0}, \beta_{0})$ used to generate the gamma data was contained in this HPD set. We implemented this process for $n = \{10^1, 10^2, 10^3, 10^4, 10^5\}$. We then repeated this process for analysis priors $p(\boldsymbol{\theta})$ that joined the marginal gamma priors from ``nature's'' prior with a Gaussian copula parameterized by Pearson's $\rho = \{0, 0.05, \dots, 0.95\}\setminus\{0.4\}$. The results from this numerical study are visualized in Figure \ref{fig:cov}.

Given the marginal $\text{GAMMA}(1000,5000)$ prior for $\alpha$ that defines $p_D(\boldsymbol{\theta})$, the posterior dependence structure between $\alpha$ and $\beta$ should converge to that of a Gaussian copula with Pearson's $\rho$ ranging between 0.86 and 0.9. We therefore expect the strength of the positive dependence between $\alpha$ and $\beta$ to increase along with the sample size $n$ for nearly all $\rho$ values for the analysis prior considered in Figure \ref{fig:cov}. As in Figure 3 of the main text, the empirical coverage in Figure \ref{fig:cov} exceeds the nominal level for small sample sizes $n$ when the strength of the dependence between the components in $\boldsymbol{\theta}$ is slightly understated. As $n$ increases, the impact of prior dependence is once again reduced and the empirical coverage approaches roughly 95\% for all $\rho$ values considered. In general, the results from this numerical study support the recommendations for prior dependence structure specification provided in Section 3.3 of the main text.


\section{Proof of Theorem 2}\label{sec:appendix.thm2} 

\textbf{Proof of Theorem 2\emph{(a)}}. The posterior mode $\tilde{\boldsymbol{\theta}}^{_{(k)}}$ minimizes $-\text{log}(p_k(\boldsymbol{\theta}\hspace{1pt}|\hspace{1pt} \boldsymbol{y}^{_{(n)}}))$ for $k = 1,2$. We have that  
\begin{equation}\label{eq:minus_p2}
\begin{split}
&-\text{log}(p_2(\boldsymbol{\theta}\hspace{1pt}|\hspace{1pt} \boldsymbol{y}^{_{(n)}})) \\ &= - l(\boldsymbol{\theta}; \hspace{1pt} \boldsymbol{y}^{_{(n)}})-\sum_{j = 1}^d \text{log}(f_j(\theta_j)) - \text{log}(c_1(\boldsymbol{u})) + (\text{log}(c_1(\boldsymbol{u})) - \text{log}(c_2(\boldsymbol{u}))) + A,
\end{split}
\end{equation}
where $l(\boldsymbol{\theta}; \hspace{1pt} \boldsymbol{y}^{_{(n)}})$ is the log-likelihood function and the constant $A$ reflects marginal likelihood term. The first three terms in (\ref{eq:minus_p2}) are equal to $-\text{log}(p_1(\boldsymbol{\theta}\hspace{1pt}|\hspace{1pt} \boldsymbol{y}^{_{(n)}}))$. When $\boldsymbol{\theta} = \tilde{\boldsymbol{\theta}}^{_{(1)}}$, it follows that $-\nabla_{\boldsymbol{\theta}} \text{log}(p_1(\boldsymbol{\theta}\hspace{1pt}|\hspace{1pt} \boldsymbol{y}^{_{(n)}})) = 0$ and $-\nabla_{\boldsymbol{\theta}}\text{log}(p_2(\boldsymbol{\theta}\hspace{1pt}|\hspace{1pt} \boldsymbol{y}^{_{(n)}})) = \nabla_{\boldsymbol{\theta}} \big[\text{log}(c_1(\boldsymbol{u})) - \text{log}(c_2(\boldsymbol{u}))\big]$. For large sample sizes $n$, the conditions for Theorem 2 ensure that the log-posterior $-\text{log}(p_2(\boldsymbol{\theta}\hspace{1pt}|\hspace{1pt} \boldsymbol{y}^{_{(n)}}))$ converges to its quadratic approximation about $\boldsymbol{\theta} = \boldsymbol{\theta}_0$. We use $\tilde{\boldsymbol{\theta}}^{_{(1)}}$ as a starting point for the Newton–Raphson method \citep{nocedal2006numerical} with $-\text{log}(p_2(\boldsymbol{\theta}\hspace{1pt}|\hspace{1pt} \boldsymbol{y}^{_{(n)}}))$ to find $\tilde{\boldsymbol{\theta}}^{_{(2)}}$. Because the quadratic approximation is appropriate for large samples, only one iteration of the Newton–Raphson method is required to approximate $\tilde{\boldsymbol{\theta}}^{_{(2)}}$. We let $\mathcal{J}(\cdot)$ be the observed information. It follows that
\begin{equation}\label{eq:newton}
\tilde{\boldsymbol{\theta}}^{_{(2)}} \approx \tilde{\boldsymbol{\theta}}^{_{(1)}} - \mathcal{J}(\tilde{\boldsymbol{\theta}}^{_{(1)}})^{-1}\nabla_{\boldsymbol{\theta} = \tilde{\boldsymbol{\theta}}^{_{(1)}}} \big[\text{log}(c_1(\boldsymbol{u})) - \text{log}(c_2(\boldsymbol{u}))\big].
\end{equation}

The conditions for Theorem 2 also guarantee that the posterior mode $\tilde{\boldsymbol{\theta}}^{_{(1)}}$ is approximately equal to the approximately normal MLE $ \hat{\boldsymbol{\theta}}^{_{(1)}}$ for large samples $\boldsymbol{y}^{_{(n)}}$. Because the MLE is consistent, $\mathcal{J}(\tilde{\boldsymbol{\theta}}^{_{(1)}})^{-1} \approx \mathcal{J}(\boldsymbol{\theta}_0)^{-1}$ for sufficiently large samples. By (8) in the main text, we have that $\nabla_{\boldsymbol{\theta} = \boldsymbol{\theta}_0} \big[\text{log}(c_1(\boldsymbol{u})) - \text{log}(c_2(\boldsymbol{u}))\big] \ne \boldsymbol{0}$, so $\text{log}(c_1(\boldsymbol{u})) - \text{log}(c_2(\boldsymbol{u}))$ can be approximated by a plane (with common gradient) in a neighbourhood of $\boldsymbol{\theta}_0$. It follows that $\mathcal{J}(\tilde{\boldsymbol{\theta}}^{_{(1)}})^{-1}\nabla_{\boldsymbol{\theta} = \tilde{\boldsymbol{\theta}}^{_{(1)}}} \big[\text{log}(c_1(\boldsymbol{u})) - \text{log}(c_2(\boldsymbol{u}))\big]$ will be roughly constant for all large samples $\boldsymbol{y}^{_{(n)}}$ generated from $m(\hspace{1pt} \cdot \hspace{1pt}| \hspace{1pt}\boldsymbol{\theta}_0)$. For an arbitrary large sample $\boldsymbol{y}^{_{(n)}}$, we have that $\tilde{\boldsymbol{\theta}}^{_{(2)}} \approx \tilde{\boldsymbol{\theta}}^{_{(1)}} + \boldsymbol{b}$ for some constant $\boldsymbol{b}$ that does not depend on $\boldsymbol{y}^{_{(n)}}$ by (\ref{eq:newton}). Since the posterior concentrates around $\boldsymbol{\theta}_0$ as the sample size $n$ increases, this common perturbation $\boldsymbol{b}$ will decrease in magnitude. $\tilde{\boldsymbol{\theta}}^{_{(1)}}$ is approximately normally distributed about $\boldsymbol{\theta}_0$ for large samples, so these small perturbations will shift $\tilde{\boldsymbol{\theta}}^{_{(2)}} \approx \tilde{\boldsymbol{\theta}}^{_{(1)}} + \boldsymbol{b}$ closer to $\boldsymbol{\theta}_0$ with probability of roughly 0.5 due to the symmetry of the normal distribution. $\hspace*{-5pt}\qed$ 

\textbf{Proof of Theorem 2\emph{(b)}}. The results from (\ref{eq:minus_p2}) and (\ref{eq:newton}) hold true as in part \emph{(a)}. Because $\boldsymbol{u}_0$ is a local minimum of $\text{log}(c_1(\boldsymbol{u})) - \text{log}(c_2(\boldsymbol{u}))$, the local linear approximation of this function is not serviceable at $\boldsymbol{\theta} = \boldsymbol{\theta}_0$. However, this implies that $- \nabla_{\boldsymbol{\theta} = \tilde{\boldsymbol{\theta}}^{_{(1)}}} \big[\text{log}(c_1(\boldsymbol{u})) - \text{log}(c_2(\boldsymbol{u}))\big]$ should be directed toward $\boldsymbol{\theta}_0$ for large samples $\boldsymbol{y}^{_{(n)}}$ when the quadratic approximation to the log-posterior is appropriate and $\tilde{\boldsymbol{\theta}}^{_{(1)}} \approx \boldsymbol{\theta}_0$. We let the orthonormal basis $\mathcal{B}$ be composed of the eigenvectors of $\mathcal{J}(\tilde{\boldsymbol{\theta}}^{_{(1)}})^{-1}$. Since $\mathcal{J}(\tilde{\boldsymbol{\theta}}^{_{(1)}})^{-1}$ is a positive definite matrix, the angle between $- \nabla_{\boldsymbol{\theta} = \tilde{\boldsymbol{\theta}}^{_{(1)}}} \big[\text{log}(c_1(\boldsymbol{u})) - \text{log}(c_2(\boldsymbol{u}))\big]$ and  $- \mathcal{J}(\tilde{\boldsymbol{\theta}}^{_{(1)}})^{-1}\nabla_{\boldsymbol{\theta} = \tilde{\boldsymbol{\theta}}^{_{(1)}}} \big[\text{log}(c_1(\boldsymbol{u})) - \text{log}(c_2(\boldsymbol{u}))\big]$ will be acute. With respect to $\mathcal{B}$, the perturbation from $\tilde{\boldsymbol{\theta}}^{_{(1)}}$ to $\tilde{\boldsymbol{\theta}}^{_{(2)}}$ induced by the Newton-Raphson method is then directed in the same orthant of $\boldsymbol{\Theta} \subseteq \mathbb{R}^{d}$ that contains $\boldsymbol{\theta}_0$. It follows that for large samples, $\tilde{\boldsymbol{\theta}}^{_{(2)}}$ cannot be further from $\boldsymbol{\theta}_0$ than $\tilde{\boldsymbol{\theta}}^{_{(1)}}$ due to a perturbation in the wrong direction. However, $\tilde{\boldsymbol{\theta}}^{_{(2)}}$ could still be further from $\boldsymbol{\theta}_0$ than $\tilde{\boldsymbol{\theta}}^{_{(1)}}$ if the magnitude of the perturbation is too large. We argue that this cannot occur for an arbitrary large sample $\boldsymbol{y}^{_{(n)}}$ because both $p_1(\boldsymbol{\theta}\hspace{1pt}|\hspace{1pt} \boldsymbol{y}^{_{(n)}})$ and $p_2(\boldsymbol{\theta}\hspace{1pt}|\hspace{1pt} \boldsymbol{y}^{_{(n)}})$ will concentrate around $\boldsymbol{\theta}_0$ and $\boldsymbol{\theta}_0$ maximizes $\text{log}(c_2(\boldsymbol{u})) - \text{log}(c_1(\boldsymbol{u}))$ in a neighbourhood of this fixed point. If this perturbation is too large for a given sample $\boldsymbol{y}^{_{(n)}}$, this behaviour cannot persist as $n \rightarrow \infty$. For large samples, these small perturbations by the Newton-Raphson method will therefore shift $\tilde{\boldsymbol{\theta}}^{_{(2)}}$ closer to $\boldsymbol{\theta}_0$ with probability approaching 1. $\hspace*{-5pt}\qed$   
	
\bibliographystyle{chicago}